\documentclass[twoside,leqno]{article}

\usepackage[letterpaper]{geometry}

\usepackage{ltexpprt}

\usepackage{amsmath}
\usepackage{amssymb}
\usepackage{braket}
\usepackage{booktabs}
\usepackage{algorithm}
\usepackage{graphicx}
\usepackage{braket}
\usepackage{enumerate}
\usepackage[colorlinks=true, allcolors=blue]{hyperref}
\usepackage{algorithm}
\usepackage[numbers]{natbib}

\begin{document}

\newcommand\relatedversion{}

\title{\Large Recovering the original simplicity: succinct and deterministic quantum algorithm for the welded tree problem}
\author{Guanzhong Li\thanks{Institute of Quantum Computing and Computer Theory, School of Computer Science and Engineering, Sun Yat-sen University, Guangzhou, China.
Supported by the National Natural Science Foundation of China Grant No. 62272492, and the Guangdong Basic and Applied Basic Research Foundation Grant No. 2020B1515020050.
Authors sorted alphabetically. Email: ligzh9@mail2.sysu.edu.cn, lilvzh@mail.sysu.edu.cn, luojq25@mail2.sysu.edu.cn}
\and Lvzhou Li\footnotemark[1]
\and Jingquan Luo\footnotemark[1]}

\date{}

\maketitle







\begin{abstract} \small\baselineskip=9pt 
This work revisits  quantum algorithms for the  well-known welded tree problem, proposing  a very succinct quantum algorithm based  on  the simplest coined quantum walks.  It  simply  iterates the naturally defined coined quantum walk operator  for a  predetermined time  and finally measure, where the predetermined time  can be efficiently computed on classical computers.  Then, the algorithm returns the correct answer deterministically, and achieves exponential speedups over any classical algorithm.  
The significance of the results may be seen as follows. (i) Our algorithm is rather simple compared with the one in (Jeffery and Zur, STOC'2023), which not only breaks the stereotype that   coined quantum walks can only achieve   quadratic speedups over  classical algorithms, but also demonstrates the power of the simplest   quantum walk model. (ii) Our algorithm theoretically achieves zero-error, which is not possible with existing methods. Thus, it becomes one of the few examples that exhibit exponential separation between  deterministic (exact) quantum and  randomized query complexities, which  may also change people's perception that since quantum mechanics is inherently probabilistic, it impossible to have a deterministic quantum algorithm with exponential speedups for the weled tree problem.
\end{abstract}

\section{Introduction}
A primary goal of the field of quantum computing is to design quantum algorithms that can solve problems faster than classical algorithms. Quantum walks have developed into a fundamental tool for algorithmic design.
Since  Aharonov et al.~\cite{AharonovDZ93PhysRevA} first coined the term ``quantum walks'' thirty years ago, quantum walks have become a major research subject both in theory and in experiment \cite{Kempe_overview, ambainis_overview, Venegas-Andraca2012, systematic}.
There are two kinds of quantum walks: discrete time quantum walks (DTQW) and continuous time quantum walks (CTQW).
Whereas  CTQW evolve a Hamiltonian $H$ (related to the graph under consideration) for any time $t$, i.e. simulating $e^{iHt}$, DTQW can only evolve the system for discrete time steps, i.e. applying $U_\mathrm{walk}^h$ to the initial state for some integer $h$ and unitary operator $U_\mathrm{walk}$.  

DTQW can be further divided into  many different frameworks. The earliest and simplest is the coined quantum walks \cite{AmbainisBNVW01,AmbainisKV01} consisting of a coin operator $C$ (usually the Grover diffusion) and a shift operator $S$ (usually the flip-flop shift, i.e. SWAP operator).
Later, Szegedy proposed a quantum walk framework \cite{Szegedy_03} from the perspective of Markov chains.
In this direction, a series of variant  frameworks for spatial search have been developed: the MNRS framework \cite{MagniezNRS11}, the interpolated walk \cite{KroviMOR16}, the electric network framework \cite{belovs2013quantum} and its finding version \cite{unified}.
Quantum algorithms based on these frameworks  have provided only at most a quadratic speedup when comparing to the best classical algorithm.
Typical examples include quantum algorithms for the element distinctness problem \cite{Ambainis07}, matrix product verification \cite{BuhrmanS06}, triangle finding \cite{MagniezSS07}, group commutativity \cite{MagniezN07}, and so on.

In sharp contrast, exponential algorithmic speedups  can be obtained based on CTQW  for the welded tree problem~\cite{CCD03}, which  makes the welded tree problem of great interest, as it is one of the few problems for which quantum walk-based algorithms are exponentially faster than classical algorithms.
Note that earlier studies have shown that quantum walks can solve problems exponentially faster than classical walks~\cite{childs2002example,kempe2002quantum}, but there exist classical  efficient algorithms  which are not based on a random walk for those problems~\cite{CCD03}.

Recently, Jeffery and Zur \cite{multi} proposed a new DTQW framework---multidimensional quantum walks (an extension of the electric network framework),  and then presented a quantum algorithm based on it to solve the welded tree problem, which  achieves exponential speedups over  classical algorithms.
The pursuit of exponential algorithmic speedups based on DTQW is one of the  reasons for proposing the framework of multidimensional quantum walks. Actually, Jeffery and Zur \cite{multi} claimed that the major drawback of the existing DTQW frameworks is that they
can achieve at most a quadratic speedup over the best classical algorithm, but this drawback does not hold for the
multidimensional quantum walk framework~\footnote{Ref.~\cite{multi} claimed ``While quantum walk frameworks
make it extremely easy to design quantum algorithms, even without an in-depth knowledge of quantum
computing, as evidenced by their wide application across domains, the major drawback is that they
can achieve at most a quadratic speedup over the best classical algorithm. This is because a quantum
walk search algorithm essentially takes a classical random walk algorithm, and produces a quantum
algorithm that is up to quadratically better. This drawback does not hold for the multidimensional quantum walk framework''.}.

In this work, we revisit quantum algorithms for the welded tree problem, proposing a quite succinct quantum algorithm based purely on the simplest coined quantum walks, which not only maintains the exponential speedup, but also achieves zero-error theoretically.
Thus,  our work demonstrates the power of the simplest  quantum walk model, and  verifies that a good quantum algorithm does not necessarily resort to complex techniques.

\subsection{Coined Quantum Walk}
For  a graph $G=(V,E)$ and $u \in V$, $deg(u)=\{v: (u,v)\in E\}$ denotes the set of neighbours of $u$, and the degree of $u$ is denoted as $d_u=|deg(u)|$.  A coined quantum walk on  $G=(V,E)$  is defined on the state space $\mathbb{H}^{N^2}= \mathrm{span}\{\ket{uv}, u,v \in V\}$ with $N=|V|$. The evolution operator of the coined quantum walk at each step is $ U_\mathrm{walk}=SC$.
 $C$,  the coin operator, is defined by 
$C=\sum_{u\in V}\ket{u}\bra{u}\otimes C_u,$
 where $C_u$ is typically the Grover diffusion coin operator $C_u=2\ket{s_u}\bra{s_u}-I,$
with $\ket{s_u}= \frac{1}{\sqrt{d_u}} \sum_{v\in deg(u)}\ket{v}.$  $S$,  the  flip-flop shift operator,  is  defined as $S\ket{uv}=\ket{vu},$  where $\ket{uv}=\ket{u} \otimes \ket{v}$  denotes a particle 
at vertex $u$ pointing towards vertex $v$.
Given the initial state $\ket{\Psi_0}$ ,  the walker's state after $h$ steps is
$\ket{\Psi_h}=U_\mathrm{walk}^h\ket{\Psi_0}$.

\subsection{The welded tree problem}\label{sec:def=tree}
\begin{figure}
	\centering
	\includegraphics[width=0.6\textwidth]{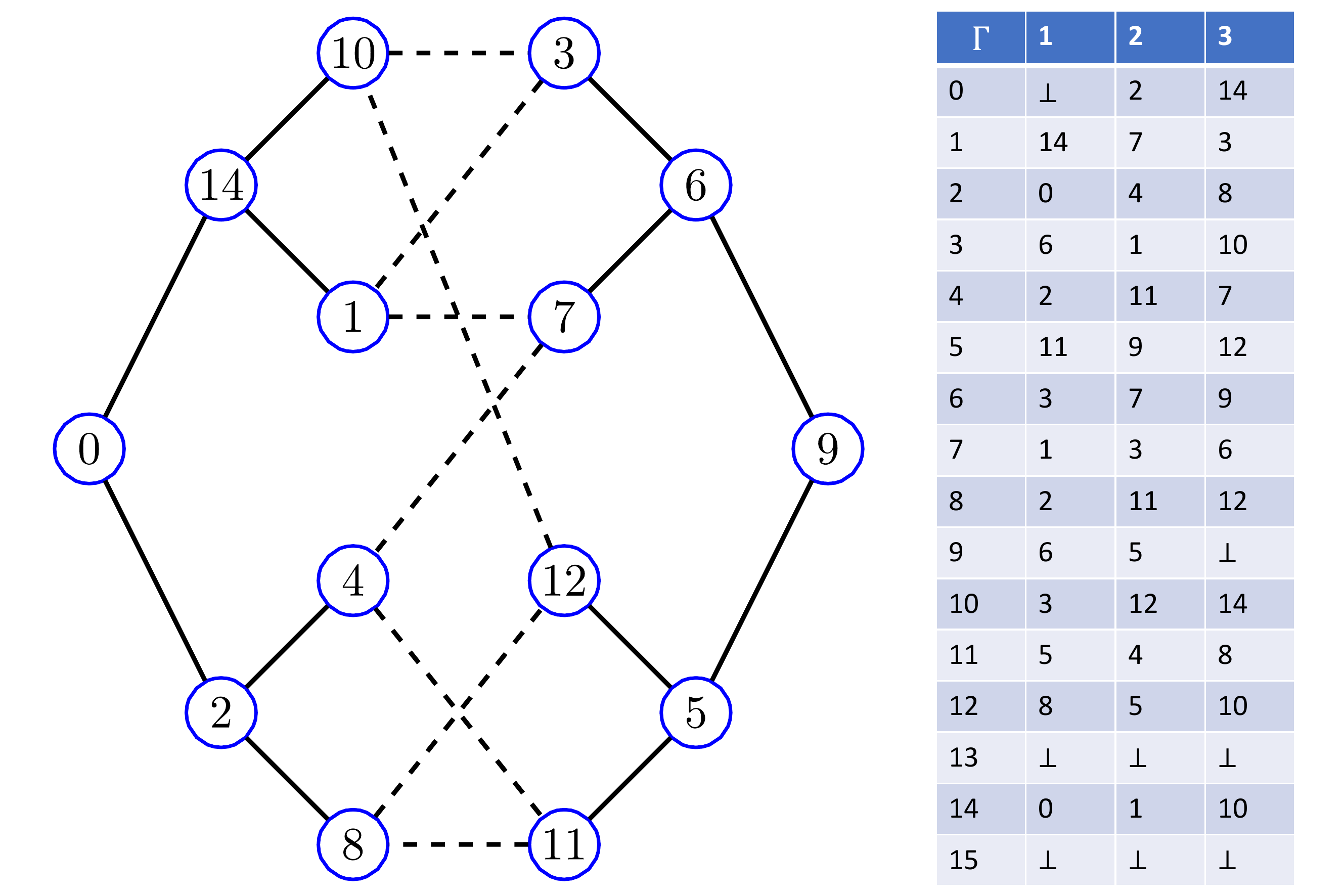}
	\caption{\label{fig:welded_graph} A welded tree $G_n$ for $n=2$ and its $2^{2n} \times 3$ adjacency list $\Gamma$. $s=0$ is the entrance and $t=9=(1001)_2$ is the exit. The dashed lines in the middle is the random cycle connecting the two trees.}
\end{figure}

The welded tree problem was proposed in \cite{CCD03} as a black-box (oracle) problem to show that a quantum algorithm can be exponentially faster than any classical algorithm with the help of CTQW. To achieve exponential \textit{algorithmic} speedup, the graph to be traversed is carefully designed:
the welded tree $G_n$ consists of two horizontal-positioned full binary trees of height $n$ with their $2^n$ leaves in the middle.
The root of the left tree is the entrance denoted by $s$, and the root of the right tree is the exit denoted by $t$. 
The leaves in the middle are connected by a \textit{random cycle} that alternates between the leaves of the two trees instead of identifying them in the direct way. (See the dashed line in Fig.~\ref{fig:welded_graph})
This makes the leaves have degree $3$ instead of $2$, indistinguishable from all the other internal nodes.

The number of vertices in $G_n$ is $|V(G_n)|=2(2^{n+1}-1)$, thus $(n+2)$-bit string is enough to encode all the vertices. However, to ensure the classical hardness of the problem~\cite[Lemma~7]{CCD03}, each vertex $u\in V(G_n)$ is \textit{randomly} assigned a distinct $2n$-bit string as its name, except that the entrance is fixed as $s \equiv 0^{2n}$.
We retain $\perp \equiv 1^{2n}$ as the special symbol, so that the graph $G_n$ can be specified by an $2^{2n} \times 3$ \textit{adjacency list} $\Gamma$ (See the table in Fig.~\ref{fig:welded_graph}).
Note that when $u\in \{0,1\}^{2n}$ is the root $s$ or $t$, exactly one of $\{\Gamma(u,i):i=1,2,3\}$ is $\perp$.

The adjacency list $\Gamma$ is provided in the form of an oracle (black box) $O$, so that the only way to know about $\Gamma$ is to query $O$ with a name $u\in \{0,1\}^{2n}$, and the oracle will output all the items in row $u$ of $\Gamma$:
\begin{equation}\label{eq:oracle_classic}
	O(u)=\{\Gamma(u,i):i=1,2,3\}.
\end{equation}
We are concerned with the number of times an algorithm queries the oracle $O$ (a.k.a query complexity) to find the exit name $t$.
In the quantum model, the effect of $O$ is
\begin{equation}\label{eq:oracle}
	O \ket{u} \bigotimes_{i=1}^{3} \ket{v_i} =
	\ket{u} \bigotimes_{i=1}^{3} \ket{v_i \oplus \Gamma(u,i)},
\end{equation}
where $u,v_i$ are all $2n$-bit string and $\oplus$ denotes bit-wise modulo $2$ addition.
The welded tree problem can now be formally stated as Definition~\ref{prob:welded}.

\begin{Definition}[the welded tree problem]
\label{prob:welded}
	Given the entrance name $s=0^{2n}$, find the exit name $t$ of the welded tree $G_n$ with as few queries as possible to its adjacency list oracle $O$.
\end{Definition}

Since the degree of each vertex in $G_n$ is no more than $3$, even if the quantum oracle $O$ is provided in its weaker form such that it returns only one adjacent vertex $\Gamma(u,i)$ when queried with $(u,i)$, as is the case in \cite{CCD03}, the influence on the query complexity is at most by a constant factor and can be neglected.

\subsection{Our contribution}
In this paper, we propose a  rather succinct quantum algorithm  (Algorithm~\ref{alg:succinct} in Section~\ref{sec:rigorous}) to solve the welded tree problem, which is simply to walk on the welded tree with  the operator $U_\mathrm{walk}=SC$ from an initial state $\ket{\Psi_s}$ that is constructed from the entrance vertex $s$. We will show that the coin operator $C$ can be implemented with $4$ queries to the given oracle $O$ and the shift operator $S$ has no queries to $O$. Also, we will prove that
after $T\in O(n \log n)$ steps which can be predetermined efficiently on classical computers, the walker reaches the exit vertex $t$ with probability at least $\Omega(\frac{1}{n})$.  More exactly, there is $|\bra{\Psi_t}U_\mathrm{walk}^T\ket{\Psi_s}|^2>c\frac{1}{n}$ for constant $c$, where $\ket{\Psi_t}$ encodes the exit $t$. Therefore, the query complexity of the algorithm is $O(n^2 \log n)$. Furthermore, the algorithm can  be improved to  a deterministic  one with $O(n^{1.5} \log n)$ queries by using deterministic (or exact) amplitude amplification as shown by Algorithm~\ref{alg:main} in Section~\ref{sec:alg}. In addition, we conjecture  that the actual complexity of our algorithm is $O(n^{4/3})$ (Conjecture~\ref{conj:walk_prob} in Section~\ref{subsec:numeri}), which is strongly supported by numerical simulation for $n=6,\cdots,500$, with the strict proof  left as an open problem.

The significance of our results, in our opinion, lies at least in the following two aspects:
\begin{enumerate}[(1)]

    \item Our algorithm is rather succinct  compared with the one in \cite{multi}, which not only changes the stereotype  that coined quantum walks can only achieve   quadratic speedups over  classical algorithms, but also demonstrates the power of the simplest  quantum walk models.

    \item Our algorithm can be made zero-error theoretically, whereas   existing methods cannot (see Table~\ref{tab:summary}).
    Thus, it becomes one of the few examples that exhibits  exponential separation between deterministic (exact) quantum and randomized query complexities, and may have potential applications in graph property testing problems \cite{childs2020can, Ben_2020}.
    Previous examples of this kind of separation include Simon's problem \cite{exact_simon} and its generalization \cite{GSP}.
    This deterministic algorithm may also change people's perception that since quantum mechanics is inherently probabilistic,  deterministic quantum algorithms with exponential speedups for the welded tree problem are out of the question.
      
\end{enumerate}

\subsection{Technical overview}\label{subsec:tech_overview}
Despite the succinctness of our algorithms, there are some non-trivial steps in designing and analyzing it, without losing technical challenges:
\begin{enumerate}[(1)]
    \item Constructing the operator $U_\mathrm{walk}=SC$ from the given oracle $O$ (Lemma~\ref{lem:ref_A}).
    As the flip-flop shift operator $S$ requires no oracle queries, the key is to construct the coin operator $C$. 
    Our implementation of $C$ is inspired by~\cite{multi}, but it is much simpler as alternative neighbourhoods technique is not needed in this paper.

    \item Reducing the $\Theta(2^n)$-dimensional state space to a $(4n+2)$-dimensional invariant subspace (Lemma~\ref{lem:subspace}).
    In this subspace, the operator $U_\mathrm{walk} = SC$ takes the form of a $(4n+2)$-dimensional square matrix  $M_U = M_S M_C$, and the initial state corresponds to the first base vector $\ket{0}$ (whose transpose is $[1,0,\cdots,0]$) and the target state corresponds to the last base vector $\ket{4n+1}$ (whose transpose is $[0,\cdots,0,1]$.
    The reduction is done by grouping the vertices according to their layers, which is inspired by~\cite{CCD03}, but since our coined quantum walk is carried out on the edges of the graph, there is some  nontrivial difference.
    \item Analyzing the success probability, which is probably the most technical step.
    This is shown in two steps:
    \begin{enumerate}[(i)]
        \item Obtaining the spectral decomposition of matrix $M_U = \sum_{j} e^{i\varphi_j} \ket{E_j} \bra{E_j}$ (Lemma~\ref{lem:spectral_decomp}). This is inspired by a spectral decomposition result in~\cite{localized}, but we improve it with an observation concerning Chebyshev polynomial of the second kind. The improvements make the equation that the eigenvalues need to satisfy become clear, explicit and easy to analyze.

        \item Instead of directly considering a fixed iteration number $t$ (the difficulty of this approach is noted in Remark~\ref{rem:fixed_t}),
        we will prove that the average success probability $\mathbb{E} |\bra{4n+1} M_U^t \ket{0}|^2$ has a $\Omega(\frac{1}{n})$ lower bound when $t$ is chosen according to a specific distribution over $\{0,1,\cdots, O(n \log n) \}$.
        A key component of the proof is the helper Lemma~\ref{lem:irwin} (inspired by~\cite[Lemma~3]{upper_improve}), which relates the lower bound of $\mathbb{E} |\bra{4n+1} M_U^t \ket{0}|^2$ with, roughly speaking, the first and last component of eigenvector $\ket{E_j}$ and the characteristic of eigenvalue angles $\varphi_j$.
        Thus, the spectral decomposition $M_U = \sum_{j} e^{i\varphi_j} \ket{E_j} \bra{E_j}$ in step (i) is of crucial importance.
        
    \end{enumerate} 
\end{enumerate}

\subsection{Related work}\label{subsec:related}
The original algorithm proposed by Childs et al.~\cite{CCD03} is based on CTQW and
they prove that the CTQW will find the exit with probability $\Omega(1/n)$ at a time of $O(n^4)$.
They also showed that the CTQW $e^{iHt}$ can be simulated for time $t$ with $O(t^2)$ oracle queries. Thus combined with fixed-point amplitude amplification \cite{QSVT,fixed_point}, the overall query complexity is $O(n^{8.5})$, where $n^{8.5}= n^{1/2} \cdot n^{4\times2} $.
Lately it was improved to $O(n^{2.5} \text{log}^2 n)$ \cite{upper_improve}.
In contrast, any classical algorithm requires $2^{\Omega(n)}$ queries \cite{CCD03,lower_improve}. It was claimed in \cite{childs2010relationship} that there exist  exponential algorithmic speedups based on DTQW  for the welded tree problem, but no  explicit algorithm was given there.
Recently, a quantum algorithm based on the multidimensional quantum walk framework was proposed by Jeffery and Zur \cite{multi}, solving the problem with $O(n)$ queries and $O(n^2)$ time complexity.
The framework uses phase estimation~\cite{phase_estimation} to gain one-bit information about the exit name, and then uses the Bernstein-Vazirani algorithm~\cite{BV} to obtain the whole name.
We summarize previous results on the welded tree problem in Table~\ref{tab:summary}.

\begin{table}[ht]
	\centering
	\begin{tabular}{cccc}
		\toprule
		algorithm type & queries & succinct? & deterministic? \\
		\midrule
		classical \cite{CCD03,lower_improve}
		& $2^{\Omega(n)}$ & 
		& * \\
		CTQW \cite{CCD03}
		& $O(n^{8.5})$ & Yes
		& No \\
		CTQW \cite{upper_improve}
		& $O(n^{2.5} \text{log}^2 n)$ & Yes
		& No \\
		DTQW \cite{multi}
		& $O(n)$ & No
		& No \\
		DTQW, this work
		& $O(n^{1.5} \log n)$ & Yes
		& Yes \\
		\bottomrule
	\end{tabular}    
	\caption{Summary of the query complexity of different algorithms for the welded tree problem. The star notation * in the first line stresses that the lower bound holds for any classical randomized algorithm and thus also holds for deterministic algorithm.
 }
	\label{tab:summary}
\end{table}

Compared to the recent quantum algorithm based on multidimensional quantum walks~\cite{multi}, our algorithm has the following advantages:
\begin{enumerate}[(1)]
    \item Succinct algorithmic procedure. Our algorithm (see Algorithm~\ref{alg:succinct}) simply iterates the coined quantum walk operator $U_\mathrm{walk}$ for a predetermined time $T$ and then measure the first register to obtain the result.
    In contrast, \cite{multi} has to combine with Bernstein-Vazirani algorithm~\cite{BV} in order to learn the whole name, because their multidimensional quantum walk framework using phase estimation can only obtain one-bit information about the exit name (which corresponds to the inner-product oracle in BV algorithm).

    \item Simpler implementation of the quantum walk operator.
    Our implementation of $U_\mathrm{walk}$ is simpler as the coin operator $C$ can be easily implemented with two oracle queries and the shift operator $S$ is merely $2n$ parallel SWAP gates.
    In contrast, it was said in \cite[Remark 4.8]{multi} that one has to carefully assign different weights to the graph's edges in order to balance between the positive and negative witness size, so that the polynomial query complexity is possible.
    The weight assigning scheme makes the implementation of the quantum walk operator much more complicated: (i) The alternative neighbourhoods technique has to be used since the oracle does not provide information about which neighbouring vertex is closer to the root. 
    The technique works as follows: instead of reflecting around the uniform superposition of neighbours (which is what our coin operator $C$ does), one has to reflect around a subspace spanned by some easily preparable states. This makes the operator's implementation more complicated. (ii) Due to the specific weight assigning scheme, one has to handle separately the cases when $n$ is odd or even, and determine the parity of the layers at which each vertex lies, and also assign different weight $w_0=w_M=1/(cn)$ to the edges $(s,\perp),(t,\perp)$.

    \item Certainty of success theoretically.
    Since the multidimensional framework~\cite{multi} uses phase estimation which is intrinsically randomized, it cannot be made deterministic, but our simple coined quantum walk algorithm can be made zero-error theoretically (Algorithm~\ref{alg:main}).
\end{enumerate}

Compared to the algorithms based on CTQW \cite{CCD03,upper_improve}, our algorithms has a better query complexity, and can be adapted to succeed with certainty. On the contrary, the implementation of the CTQW operator $e^{iHt}$ from oracle $O$ involves the use of linear combination tool in Hamiltonian simulation, thus error is introduced inevitably. Also, choosing the quantum walk time $t$  according to some distribution leads to additional randomness.

\subsection{Paper organization}
The rest of the paper is organized as follows.
In Section~\ref{sec:construct} we define the coined quantum walk operator $U_\mathrm{walk}$ and give a detailed implementation of  $U_\mathrm{walk}$ from the quantum oracle $O$.
In Section~\ref{sec:subspace} we reduce the full state space to a $(4n+2)$-dimensional invariant subspace and deduce the reduced matrix $M_U$, which lays an important first step for the  correctness and complexity analysis of our algorithms.
In Section~\ref{sec:rigorous} we present a succinct algorithm (Algorithm~\ref{alg:succinct}) and prove a rigorous query upper bound.
In Section~\ref{sec:alg} we present the theoretically zero-error algorithm (Algorithm~\ref{alg:main}).
Numerical simulation is shown in Section~\ref{subsec:numeri} indicating that the actual performance of our algorithms is better.
We conclude this paper in Section~\ref{sec:conclu}.

\section{Implementing the coined quantum walk operator}\label{sec:construct}
As the adjacency list $\Gamma$ of the welded tree  $G_n$  defined in Section~\ref{sec:def=tree} is unknown and can only be accessed through the oracle $O$, the implementation of the coined quantum walk operator $U_\mathrm{walk}$ on $G_n$ needs some careful design as presented in this section.

First, 
the state space of the coined quantum walk is:
\begin{equation}
	\mathcal{H}= \text{span} \{\ket{u}_{r_1} \ket{v}_{r_2} : u,v\in\{0,1\}^{2n} \},
\end{equation}
which is the state space of $4n$ qubits. The subscript $r_i$ of the two registers will serve later for the convenience of describing the construction of \begin{align}
 U_\mathrm{walk}=SC.  \label{eq:U_AB_def}
\end{align}
 Let $\varphi(u)=\frac{1}{\sqrt{3}}\sum_{i=1}^{3} \ket{\Gamma(u,i)}$ be the uniform superposition of the adjacent vertices of $u$ and let $C_u=2\ket{\varphi(u)}\bra{\varphi(u)}-I$. Then the coin operator $C$ is given by 
\begin{align}
C&= \sum_{u\in \{0,1\}^{2n}}\ket{u}\bra{u}\otimes C_u\\
 &= 2\sum_{u\in \{0,1\}^{2n} } \ket{u,\varphi(u)} \bra{u,\varphi(u)}-I.\label{eq:ref_A_def}
\end{align}
Note that we allow the sum to include $u\notin V(G_n)$, so that the implementation of $\text{Ref}_{\perp}$ (see Eq.~\eqref{eq:ref_LA_def}) does not need to check whether $u$ is indeed a vertex in $G_n$ or not.
When $r\in \{s,t\}$ and $\Gamma(r,i_1)=\perp$, we let
\begin{equation}\label{eq:varphi_r}
    \varphi(r)=\frac{1}{\sqrt{2}} (\ket{\Gamma(r,i_2)}+\ket{\Gamma(r,i_3)}),
\end{equation}
reflecting the fact that $r \to \perp$ is not an edge in the graph.

The shift operator $S$ is the SWAP operator on the vertex pair $\{u,v\}$:
\begin{equation}
S = \sum_{u,v\in \{0,1\}^{2n} } \ket{v,u}\bra{u,v},  \label{eq:ref_B_def}
\end{equation}
which is actually a reflection operator as well:
\begin{equation}
    S = 2\sum_{u \leq v} \ket{\psi_{u,v}} \bra{\psi_{u,v}}-I,
\end{equation}
where
\begin{equation}
    \ket{\psi_{u,v}} =
\begin{cases}
\frac{\ket{u,v}+\ket{v,u}}{\sqrt{2}}, & u<v; \\
\ket{u,v}, & u=v.
\end{cases}
\end{equation}

\begin{Lemma}\label{lem:ref_A}
	The coined quantum walk operator $U_\mathrm{walk}= SC$ can be implemented with $2$ oracle queries and $O(n)$ elementary operations.
\end{Lemma}
\begin{proof}
First, the implementation of $S$ is quite simple, just apply the SWAP gate to the corresponding $2n$ pairs of qubits between registers $r_1$ and $r_2$, \textit{which takes $O(n)$ basic operations}.
The implementation of $C$ is more complicated and requires $2$ oracle queries as shown below.

	We implement $C$ in two steps: first to construct a unitary operator $U_\varphi$ that has the following effect
    \begin{equation}
        U_\varphi: \ket{u,\perp} \mapsto \ket{u, \varphi(u)},
    \end{equation}
    and then to construct the reflection 
	\begin{equation}\label{eq:ref_LA_def}
		\text{Ref}_{\perp} = 2 \sum_{u \in \{0,1\}^{2n}} \ket{u,\perp} \bra{u,\perp} -I
        = I_{r_1} \otimes (2\ket{\perp}\bra{\perp}-I_{r_2}).
	\end{equation}
	Thus, we have \begin{align}
	 C = U_\varphi\,\text{Ref}_{\perp} U_\varphi^\dagger,   
	\end{align} where $U_\varphi^\dagger$ can be implemented by executing the conjugate of quantum gates composing $U_\varphi$ in reverse order.

    As global phase shift can be neglected, we will implement $-\text{Ref}_{\perp}$. Observe that $-\text{Ref}_{\perp}$ simply adds a relative phase shift of $(-1)$ to $\ket{v}_{r_2}$ when $v=\perp=1^{2n}$.
    Thus, using phase kick-back effect,  $-\text{Ref}_{\perp}$ can be easily constructed by flipping an auxiliary qubit register $\ket{-}$ conditioned on all the $2n$ qubits in register $r_2$ being in state $\ket{1}$ (i.e. apply a $\text{C}^{2n}-\text{NOT}$ gate, \textit{which decomposes to $O(n)$ basic gates}).

    The implementation of $U_\varphi$ requires $2$ oracle queries and \textit{$O(n)$ basic operations}, which is a bit lengthy due to the special handling of $\varphi(r)$ (Eq.~\eqref{eq:varphi_r}) and is thus deferred to Appendix~\ref{sec:coin_imp}.
    
    \textit{As can be seen form the italics, the implementation of $U_\mathrm{walk}$ takes $O(n)$ basic operations in total.}
\end{proof}

\begin{Remark}
If we enable the oracle to return all the neighbours coherently, i.e. $O:\ket{u,\perp} \mapsto \ket{u,\varphi(u)}$, which is a common assumption in the Markov chain based DTQW framework, then the implementation of $U_\varphi$ shown above is unnecessary.
But if this is the case, we will need an additional oracle to check if $u$ is the exit.
\end{Remark}

The initial state of the coined quantum walk is
\begin{equation}\label{eq:initial}
    \ket{s,\varphi(s)} = \frac{1}{\sqrt{2}} (\ket{\Gamma(s,i_2)}+\ket{\Gamma(s,i_3)}),
\end{equation}
which can be obtained with $2$ oracle queries from $\ket{s,\perp}$ similar to step 4 in Appendix~\ref{sec:coin_imp}. We denote this state preparation unitary by
\begin{equation}
    U_p: \ket{s,\perp} \mapsto \ket{s,\varphi(s)}.
\end{equation}

\section{Reducing to the low-dimensional invariant subspace}\label{sec:subspace}
In this section, we will determine the $(4n+2)$-dimensional invariant subspace $\mathcal{H}_0$ of the coined quantum walk operator $U_\mathrm{walk}$ based on layers of vertices in $G_n$, so that the amplitude on the target state $\ket{t,\varphi(t)}$ after applying $U_\mathrm{walk}^T$ to the initial state $\ket{s,\varphi(s)}$ can be calculated exactly when $n$ is fixed, regardless of the vertices' random naming or the random cycle in the middle of $G_n$. This  lays an important first step for the  correctness and complexity analysis of our algorithms.

Speciffically, we have the following lemma.
\begin{Lemma}\label{lem:subspace}
    The coined quantum walk operator $U_\mathrm{walk}$ for  the welded tree $G_n$ has a $(4n+2)$-dimensional invariant subspace
    \begin{equation}
        \mathcal{H}_0 = \mathrm{span} \{ \ket{0,R}, \ket{1,L}, \ket{1,R},\cdots, \ket{2n,L},\ket{2n,R}, \ket{2n+1,L} \},
    \end{equation}
    where $\ket{0,R} = \ket{s,\varphi(s)}$ is the initial state, $\ket{2n+1,L} = \ket{t,\varphi(t)}$ is the target state, and the other sates $\ket{k,L}, \ket{k,R} $ are defined respectively in Eqs.~\eqref{eq:k_L_left}~\eqref{eq:k_R_left} for $k=1\sim n$ in the left tree, and similarly for $k= (n+1) \sim 2n$ in the right tree.
    In this basis, the coined quantum walk operator $U_\mathrm{walk}$ can be represented by a  $(4n+2)$-dimensional square matrix $M_U = M_S \cdot M_C$, where $M_C$ and $M_S$ are shown in Eqs.~\eqref{eq:ref_A_m},~\eqref{eq:ref_B_m} respectively.
\end{Lemma}

\begin{proof}
The welded tree $G_n$ has $2(n+1)$ layers of vertices, and we denote by $V_k$ the set of vertices in the $k$-th layer. Thus in the left tree, $|V_k|=2^k$ for $k\in \{0,1,\cdots,n\}$; and in the right tree, $|V_{n+k}|=2^{n+1-k}$ for $k\in \{ 1,2,\cdots,n+1 \}$.
The two basis states $\ket{k,L},\ket{k,R}$ with $k\in \{1,\cdots,n\}$ of $\mathcal{H}_0$ are related to $V_k$ in the left tree, and are defined as the superpositions of the directed edges pointing to the root and to the random cycle, respectively:
\begin{equation}\label{eq:k_L_left}
	\ket{k,L} := \frac{1}{\sqrt{2^k}} \sum_{u\in V_k} \ket{u,\Gamma(u,i_1)}
\end{equation}
where $\Gamma(u,i_1)$ is the adjacent vertex of $u$ closest to the root $s$, and 
\begin{equation}\label{eq:k_R_left}
	\ket{k,R} := \frac{1}{\sqrt{2^{k}}} \sum_{u\in V_k} \frac{1}{\sqrt{2}} \big( \ket{u,\Gamma(u,i_2)} + \ket{\Gamma(u,i_3)} \big).
\end{equation}
It's easy to see that they are orthogonal, since the composing computational basis states represent distinct directed edges in the welded tree graph.
The two basis states $\ket{n+k,L}, \ket{n+k,R}$ with $k\in\{1, \cdots,n\}$ related to $V_{n+k}$ in the right tree are defined similarly.
Note that $\ket{0,R} := \ket{s,\varphi(s)}$ is the initial state and $\ket{2n+1,L} := \ket{t,\varphi(t)}$ is the target state. Note also that there is no $\ket{0,R}$ or $\ket{2n+1,L}$.
A diagram of the $4n+4=12$ basis states of $\mathcal{H}_0$ when $n=2$ is shown in Fig.~\ref{fig:H_0}.

\begin{figure}
	\centering
	\includegraphics[width=0.4\textwidth]{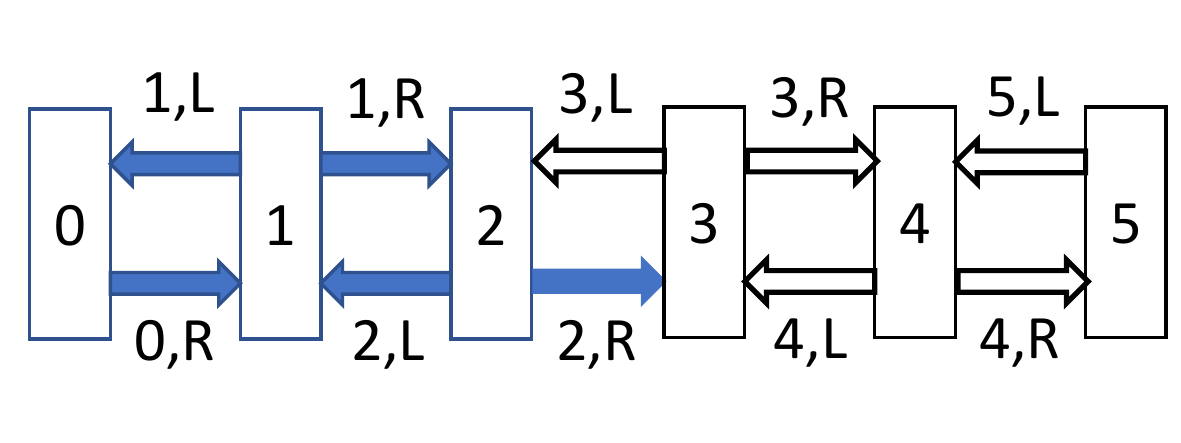}
	\caption{\label{fig:H_0} Diagram of the basis states of the reduced coined quantum walk subspace $\mathcal{H}_0$ when $n=2$. The box with number inside represents the vertex set $V_k$ of the $k$-th layer. The solid blue arrows represent basis states $\ket{k,L},\ket{k,R}$ related to $V_k$ with $k\in \{1,\cdots,n\}$ in the left tree, and the black hollow arrows represent basis states $\ket{n+k,L}, \ket{n+k,R}$ related to $V_{n+k}$ with $k\in \{ 1,2,\cdots,n \}$ in the right tree. The state $\ket{0,R}:=\ket{s,\varphi(s)}$ and $\ket{2n+1,L}:=\ket{t,\varphi(t)}$ are the initial and target states respectively. }
\end{figure}

Observe that
\begin{equation}
	\ket{u,\varphi(u)} = \sqrt{\frac{1}{3}} \ket{u,\Gamma(u,i_1)} +  \sqrt{\frac{2}{3}} \frac{1}{\sqrt{2}} \Big( \ket{u,\Gamma(u,i_2)} + \ket{\Gamma(u,i_3)} \Big).
\end{equation}
Thus by the definition of the coin operator $C$ (Eq.~\eqref{eq:ref_A_def}) and linearity, the $2$-dimensional subspace spanned by $\{\ket{k,L},\ket{k,R}\}$ is invariant under $C$, and the matrix expression of $C$ in this basis is
\begin{equation}\label{eq:R_A}
	R_A :=
	2\begin{bmatrix} \sqrt{\frac{1}{3}} \\ \sqrt{\frac{2}{3}} \end{bmatrix} \cdot
	[\sqrt{\frac{1}{3}} , \sqrt{\frac{2}{3}}] - I
	=
	\begin{bmatrix}
		-\frac{1}{3} & \frac{2\sqrt{2}}{3}\\
		\frac{2\sqrt{2}}{3} & \frac{1}{3}
	\end{bmatrix}.
\end{equation}
Similarly, the matrix expression of $C$ in the basis $\{\ket{n+k,L},\ket{n+k,R}\}$ related to $V_{n+k}$ in the right tree is
\begin{equation} \label{eq:R_A_i}
	R_A':=
	2\begin{bmatrix} \sqrt{\frac{2}{3}} \\ \sqrt{\frac{1}{3}} \end{bmatrix} \cdot
	[\sqrt{\frac{2}{3}} , \sqrt{\frac{1}{3}}] - I
	=
	\begin{bmatrix}
		\frac{1}{3} & \frac{2\sqrt{2}}{3}\\
		\frac{2\sqrt{2}}{3} & -\frac{1}{3}
	\end{bmatrix}.
\end{equation}
Note that $\ket{0,R}$ and $\ket{2n+1,L}$ are invariant under $C$ by Eq.~\eqref{eq:varphi_r}. Thus the matrix expression of $C$ in the basis  $ \left\{ \ket{0,R}, \ket{1,L}, \ket{1,R},\cdots, \ket{2n,L},\ket{2n,R}, \ket{2n+1,L} \right\}$ is
\begin{equation} \label{eq:ref_A_m}
	M_C := \text{diag}(1,\underbrace{R_A,\cdots,R_A}_{n},\underbrace{R_A',\cdots,R_A'}_{n},1).
\end{equation}

Note that $\ket{k,R}$ and $\ket{k+1,L}$ with $k\in\{0,\cdots,2n\}$ are equal superpositions of basis states in $\{ \ket{u,v}:u\in V_{k}, v\in V_{k+1} \}$ and $\{ \ket{v,u}:u\in V_{k}, v\in V_{k+1} \}$ respectively. Therefore, $S$ simply swaps $\ket{k,R}$ and $\ket{k+1,L}$, and the matrix expression of $S$ in this two basis states is
\begin{equation} \label{eq:ref_B}
	R_B := \begin{bmatrix}0 & 1 \\ 1 & 0 \end{bmatrix}=2\ket{+} \bra{+} -I.
\end{equation}
Thus, the matrix of $S$ in the basis $ \left\{ \ket{0,R}, \ket{1,L}, \ket{1,R},\cdots, \ket{2n,L},\ket{2n,R}, \ket{2n+1,L} \right\}$ is
\begin{equation} \label{eq:ref_B_m}
	M_S := \text{diag}(\underbrace{R_B,\cdots,R_B}_{2n+1}).
\end{equation}
As a result, the operator $U_\mathrm{walk}$ corresponds to a $(4n+2)$-dimensional square matrix
\begin{align}\label{eq:M_U}
    M_U=M_S\cdot M_C.
\end{align}

In addition, the initial state $\ket{0,R} := \ket{s,\varphi(s)}$ corresponds to the $(4n+2)$-dimensional vector $\ket{0}=[1,0,\cdots,0]^{\text{T}}$, and 
the target state $\ket{2n+1,L} := \ket{t,\varphi(t)}$ corresponds to $\ket{4n+1}=[0,0,\cdots,1]^{\text{T}}$.
\end{proof}


\section{Succinct quantum algorithm with $O(n^2\log n)$ queries}\label{sec:rigorous}
\begin{algorithm}[htp]
	\caption{succinct quantum algorithm for the welded tree problem}
	\label{alg:succinct}
	\begin{description}
		\item[Input:] adjacency list quantum oracle $O$ (see Eq.~\eqref{eq:oracle}) for the welded tree $G_n$, and the entrance name $s \equiv 0^{2n}$.
		\item[Output:] the exit name $t$.
		\item[Procedure:]
	\end{description}
	\begin{enumerate}
		\item Classical predetermine the walk step number $T_1$:  For the initial vector $\ket{\psi_0} = \ket{0} $, loop ``$\ket{\psi_T} \leftarrow M_U\ket{\psi_{T-1}}$'' (see Eq.~\eqref{eq:M_U} for $M_U$) and ``$ p_T \leftarrow |\braket{4n+1 | \psi_T}|$'', 
		stop when $T>3.6\,n \log(5n)$. Record the \textit{largest} $p_{T_1}$ for $2n < T_1 < 3.6\, n \log(5n)$ and the corresponding $T_1$.

        \item Quantum walk: Apply $T_1$ steps of coined quantum walk, i.e. $U_\mathrm{walk}^{T_1}$ (Eq.~\eqref{eq:U_AB_def}) to the initial state $\ket{s,\varphi(s)}$ (Eq.~\eqref{eq:initial}) and measure the first register in the computational basis to obtain a vertex name with $\Omega(\frac{1}{n})$ probability.

        \item Repeat step 2 for $O(n)$ times to obtain the exit name $t$ with constant probability.
	\end{enumerate}
\end{algorithm}
Here we present a rather succinct quantum algorithm (i.e., Algorithm \ref{alg:succinct}) for the welded tree problem, which is  to  first efficiently compute a walk step number $T_1$ on classical computers, and then simply perform the walk operator  $U_\mathrm{walk}$ with $T_1$ times. The correctness  is guaranteed by  Theorem~\ref{thm:main}, from which together with Lemma \ref{lem:ref_A} it is easy to see that the query complexity of Algorithm~\ref{alg:succinct} is $O(n^2 \log n)$, with an additional time complexity  of  $O(n\cdot n^2\log(n)) = O(n^3 \log(n))$. 

\begin{theorem}\label{thm:main}
    Consider the success probability $p(t) := \left|\bra{4n+1} M_U^t \ket{0}\right|^2$, where $M_U$ is defined by Eq.~\eqref{eq:M_U}.  Then for sufficiently large $n$ and $T\approx 3.6\, n \log(5n)$, we have
    \begin{equation}
        \max \{ p(t): t\in[2n,T]\} > \frac{1}{20n}.
    \end{equation} 
\end{theorem}
The remainder of this section is to prove the above theorem. 
As mentioned in Section~\ref{subsec:tech_overview}, the first and most important step to prove Theorem~\ref{thm:main} is to obtain the spectral decomposition of the reduced matrix, i.e. $M_U = \sum_{j} e^{i\varphi_j} \ket{E_j} \bra{E_j}$, which will be shown by Lemma~\ref{lem:spectral_decomp} in Section \ref{sec:spectral-dec}.
Then we will present and prove the helper Lemma~\ref{lem:irwin} in Section~\ref{subsec:helper_lem}, a key step in obtaining the  lower bound $\Omega(\frac{1}{n})$ of the average success probability $\mathbb{E} |\bra{4n+1} M_U^t \ket{0}|^2$, from which Theorem \ref{thm:main} follows.
In order to use this helper lemma, we will need the values $\braket{4n+1|E_j} \cdot \braket{E_j|0}$ for $j\in S$,
and bound the gap $\Delta E_S = \min_{} \{ |\varphi_j -\varphi_k|: \varphi_j\in S,\varphi_k\in E,k\neq j \}$ by choosing $S\subseteq E=\{\varphi_j\}$ well (so that $\Delta E_S$ will be of order $\Omega(\frac{1}{n})$ as shown by Eq.~\eqref{eq:relation_theta_phi} and Lemma~\ref{lem:phase_gap} in Section~\ref{subsec:delta_E}).
Thus, even though the explicit formula shown in Lemma~\ref{lem:spectral_decomp} is quite complicated, we only need to pay attention to the the first and last term of the eigenvectors $\ket{E_j}$, and the gap between the eigenvalue angles $\varphi_j$.

\subsection{Spectral decomposition of $M_U$}\label{sec:spectral-dec}

As a preliminary, first notice that $M_C$ (see Eq.~\eqref{eq:ref_A_m}) can be expressed as $2A A^\dagger -I$, where $A$ is the following $2(2n+1)\times 2(n+1)$ centrosymmetric matrix:
\begin{equation}\label{eq:matrix_A}
    A=
\left[\begin{array}{cccc|c}
1 & & & & \\
 & \sqrt{p} & & & \\
 & \sqrt{q} & & & \\
 & & \ddots & & \\
 & & & \sqrt{p} & \\
 & & & \sqrt{q} & \\
 \hline
 & & & & * \\
\end{array}\right],
\end{equation}
where
\begin{equation}
    p=\frac{1}{3},\ q=1-p
\end{equation}
represent respectively the probability of walking to the roots and to the random cycle.
The $(2n+1) \times (n+1)$ sub-matrix in the bottom right-hand corner denoted by `$*$' can be deduced from the centrosymmetry of $A$.
It's easy to see that all the columns in $A$ are orthonormal, thus $A^\dagger A =I_{2(n+1)}$.

\begin{Lemma}\label{lem:spectral_decomp}
    The matrix $M_U$ defined by Eq.~\eqref{eq:M_U} has $(4n+2)$ different eigenvalues. Two of which are $\pm 1$, and the respective eigenvectors are $\ket{u_{\pm 1}} = A \ket{v_{\pm 1}}$, where the $i$-th component of $\ket{v_{\pm 1}}$ denoted by $v_{\pm 1}(i)$ is shown in the following:
    \begin{equation}\label{eq:v_1_element}
        v_{\pm 1}(i)=
    \begin{cases}
        1, &i=1, \\
        (\pm \sqrt{q/p})^{i-1} /\sqrt{q}, &i=2\sim n+1, \\
        \pm (*) &i=n+2 \sim 2n+2.
    \end{cases}
    \end{equation}
    The $(*)$ above can be deduced from centrosymmetry of $\ket{v_{\pm 1}}$. The square of norms are:
    \begin{equation}\label{eq:v_1_norm}
        \| \ket{u_{\pm 1}} \|^2 = \| \ket{v_{\pm 1}} \|^2 = \frac{2}{p-q} \{ 2p-(q/p)^{n} \}.
    \end{equation}
    
    The other $4n$ eigenvalues are $\exp(\pm i \varphi_{\pm k})$ with $ \ k=2\sim n+1$, where $\varphi_{\pm k} = \arccos \lambda_{\pm k}$ and $\lambda_{-k}  =-\lambda_k$.
    Here, $\lambda_{\pm k}= 2\sqrt{pq} \cos\theta_{\pm k}$, and $\theta_{\pm k}$ ($\theta_{-k} :=\pi - \theta_k$) are the $2n$ roots of the following equation:
    \begin{equation}\label{eq:eigval_eqn_2}
        \sqrt{q} \sin(n+1)\theta  \pm \sqrt{p} \sin n\theta =0.
    \end{equation}
    The eigenvectors corresponding to $\exp(\pm i \varphi_{\pm k})$ are:
    \begin{equation}
        \ket{u_{\pm, \pm k}} := \ket{a_{\pm k}} - \exp(\pm i\varphi_{\pm k}) \ket{b_{\pm k}},
    \end{equation}
    where $\ket{a_{\pm k}}=A\ket{v_{\pm k}}, \ket{b_{\pm k}} = M_S\ket{a_{\pm k}}$. The components of $\ket{v_{\pm k}}$ are as follows:
    \begin{equation}\label{eq:v_k_element}
        v_{\pm k}(i) = 
        \begin{cases}
            1, &i=1, \\
            \frac{\lambda_{\pm k}} {\sqrt{p}} {U}_{i-2}(\lambda_{\pm k}/\sqrt{pq}) 
            -\frac{1}{\sqrt{q}} {U}_{i-3}(\lambda_{\pm k}/\sqrt{pq}), &i=2\sim n+1 \\
            \pm (*), &i=n+2\sim 2n+2
        \end{cases}
    \end{equation}
    where $(*)$ can be deduced from centrosymmetry, and $U_i(x)$ is the monic Chebyshev of the second kind:
    \begin{equation}\label{eq:U_k_cheby_lem}
        U_i(x) =\frac{\sin(i+1)\arccos \frac{x}{2}}{\sqrt{1-(\frac{x}{2})^2}}.
    \end{equation}
    The square of norms are
    \begin{equation}\label{eq:v_k_norm}
        \| \ket{u_{\pm,\pm k}} \|^2 =\frac{2(1-\lambda_k^2)^2}{q\sin^2\theta_k} \Big( n +\sqrt{\frac{q}{p}} \frac{\sin((n+1)2\theta_k)}{2\sin\theta_k} \Big).
    \end{equation}
\end{Lemma}

\begin{proof}
    The proof is a bit lengthy and is deferred to Appendix~\ref{sec:spectral}.
\end{proof}

\begin{Remark}
We suspect the reason that no succinct algorithm for the welded tree problem based on simple coined quantum walks has been proposed a priori, is that an enough understanding on the spectral decomposition of the DTQW operator has not been obtained before.
Although Ref.~\cite{localized} is a big step towards  this direction, the results obtained there are not satisfactory:
\begin{equation}\label{eq:unsatisfactory}
    \frac{1}{T} \sum_{t\in [T]} |\bra{4n+1} M_U^t \ket{0}|^2 =2^{-\Omega(n)}, \quad T\to \infty,
\end{equation}
showing that the average success probability is exponentially small when $T\to \infty$.
\end{Remark}

\subsection{The helper lemma}\label{subsec:helper_lem}
Another key component in proving Theorem~\ref{thm:main} is the following helper Lemma~\ref{lem:irwin}, which is a discrete-time adaptation of \cite[Lemma~3]{upper_improve}. 
However, our proof of Lemma~\ref{lem:irwin} is simpler than the one for \cite[Lemma~3]{upper_improve}, as it does not involve integral or characteristic functions of continuous random variables. Moreover, their intermediate step \cite[Lemma~4]{upper_improve} considers the Frobenius norm of the difference between density matrices, which introduces a strange factor of $\sqrt{3}$.

Lemma~\ref{lem:irwin} shows that when the iteration number $t$ is chosen according to some specific distribution on $[T] := \{0,1,\cdots,T-1\}$, the average success probability has a lower bound that is related to the characteristic of the eigenvalue angles of $M_U$ and the products of the first and the last term of the eigenvectors.

\begin{Lemma}\label{lem:irwin}
    Assume that  $M_U$ has a spectral decomposition $M_U= \sum_{j} e^{i\varphi_j} \ket{E_j} \bra{E_j}$ where $\varphi_j$ are all distinct, and the initial state is written in this eigenbasis as $\ket{\psi_0}=\sum_{j} c_j \ket{E_j}$ and the target state as $\ket{y}=\sum_j y_j \ket{E_j}$.
    For a subset $S \subseteq E:=\{ \varphi_j \}$,
    denote $\Delta E_S := \min_{} \{ |\varphi_j -\varphi_k|: \varphi_j\in S,\varphi_k\in E,k\neq j \}$. 
    Let $t=\sum_{m=1}^k t_m$ be the sum of $k$ i.i.d. uniform random variables $t_m \in [T]$. Consider the average success probability
    \begin{equation}\label{eq:aver_pr_def}
        \bar{p}(y|\psi_0) =\frac{1}{T^k} \sum_{t\in [T]^k} |\bra{y} M_U^t \ket{\psi_0}|^2,
    \end{equation}
    where  $t$ can be regarded as a random vector $(t_1,t_2,\cdots,t_k) \in [T]^k$ each with equal probability $\frac{1}{T^k}$. Then $\bar{p}(y|\psi_0)$ has the following lower bound:
    \begin{equation}\label{eq:aver_pr_lower}
        \bar{p}(y|\psi_0) \geq \sum_{j: \varphi_j \in S} |y_j^*\, c_j|^2 -\Big( \frac{\pi}{T\, \Delta E_S} \Big)^k.
    \end{equation}
\end{Lemma}


\begin{proof}
    Denote by $ A(t) := \bra{y} M_U^t \ket{\psi_0} $ the amplitude of success, then according to the spectral decomposition of $M_U$, we have $A(t) =\sum_{j} y_j^* \, c_j \, e^{i\varphi_j t}$. From $|A(t)|^2 =A(t) A(t)^*$, we know
    \begin{equation}
        \bar{p}(y|\psi_0) =\sum_{j,j'} \left[ y_j^*\, c_j\, y_{j'}\, c_{j'}^*\,
        \sum_{t\in [T]^k} \frac{1}{T^k} e^{i(\varphi_{j}-\varphi_{j'})t} \right]. \label{eq:32}
    \end{equation}
    We now divide the sum $\sum_{j,j'}[\cdots]$ in Eq.~(\ref{eq:32}) into the following three parts.
    \begin{enumerate} [(1)]
        \item $j=j'$ and $\varphi_j \in S$: $\sum_{j,j'}[\cdots] = \sum_{j\in S} |y_j^* \, c_j |^2$.
        \item $\varphi_j \notin S$ and $\varphi_{j'} \notin S$:
        \begin{equation}
            \sum_{j,j'} [\cdots] =\frac{1}{T^k} \sum_{t} \sum_j y_j^* \, c_j e^{i\varphi_j t} \sum_{j'} y_{j'} c_{j'}^* e^{-i\varphi_{j'} t} \geq 0.
        \end{equation}
        \item The rest part, i.e. $j\neq j', \varphi_j\in S, \varphi_{j'} \in S$, and $\varphi_j \notin S, \varphi_{j'} \in S$, and $\varphi_j \in S, \varphi_{j'} \notin S$.
    \end{enumerate}
    Therefore, we only need to prove that the value of part (3) is greater  than  $-\Big( \frac{\pi}{T\, \Delta E_S} \Big)^k$.
    First, note that now $\varphi_j \neq \varphi_{j'}$ and one of them belongs to $S$. Thus $|\varphi_j -\varphi_{j'}| \geq \Delta E_S$, and we have
    \begin{align}
        \Big| \sum_{t\in [T]^k} \frac{1}{T^k} e^{i(\varphi_{j}-\varphi_{j'})t} \Big|
        &= \Big| \prod_{m=1}^k  \sum_{t_m=0}^{T-1} \frac{1}{T} e^{i(\varphi_{j}-\varphi_{j'})t_m} \Big| \\
        &= \prod_{m=1}^k \Big| \sum_{t_m=0}^{T-1} \frac{1}{T} e^{i(\varphi_{j}-\varphi_{j'})t_m} \Big| \\
        &= \Big| \frac{1-e^{i(\varphi_{j}-\varphi_{j'})T}}{T(1-e^{i(\varphi_{j}-\varphi_{j'})})} \Big|^k \\
        &\leq \Big( \frac{2}{T\cdot {2} \Delta E_S / {\pi} } \Big)^k = \Big( \frac{\pi}{T \Delta E_S} \Big)^k,
    \end{align}
    where we have used $t=\sum_{m=1}^k t_m$ in the first equality, and the following identities $|1 -e^{i\varphi}| =|e^{i\varphi/2} -e^{-i\varphi/2}| =|2\sin\varphi/2| \geq 2\frac{\varphi}{\pi}$ in the last line.
    Then, we can bound the value of part (3) using Cauchy-Schwartz as follows:
    \begin{align}
        \Big| \sum_{j,j'} [\cdots] \Big| &\leq \Big( \frac{\pi}{T \Delta E_S} \Big)^k
        \sum_{j,j'} | y_j^*\, c_j\, y_{j'}\, c_{j'}^* | \\
        &\leq \Big( \frac{\pi}{T \Delta E_S} \Big)^k \sqrt{ \sum_{j,j'} |y_j|^2 |y_{j'}|^2 } \sqrt{\sum_{j,j'} |c_j|^2 |c_{j'}|^2 } \\
        &= \Big( \frac{\pi}{T \Delta E_S} \Big)^k.
    \end{align}
\end{proof}

\subsection{Lower bounding $\Delta E_S$}\label{subsec:delta_E}

With Lemma~\ref{lem:irwin} in hand, in order to prove Theorem~\ref{thm:main}, we will need to select a suitable subset of eigenvalue angles $S$ and an upper bound of iteration times $T'=k(T-1) <kT$ such that the first term of Eq.~\eqref{eq:aver_pr_lower} is $\Omega(\frac{1}{n})$, and the second term is a smaller $O(\frac{1}{n})$ term.

\begin{figure}[htbp]
	\centering
	\includegraphics[width=0.5\textwidth]{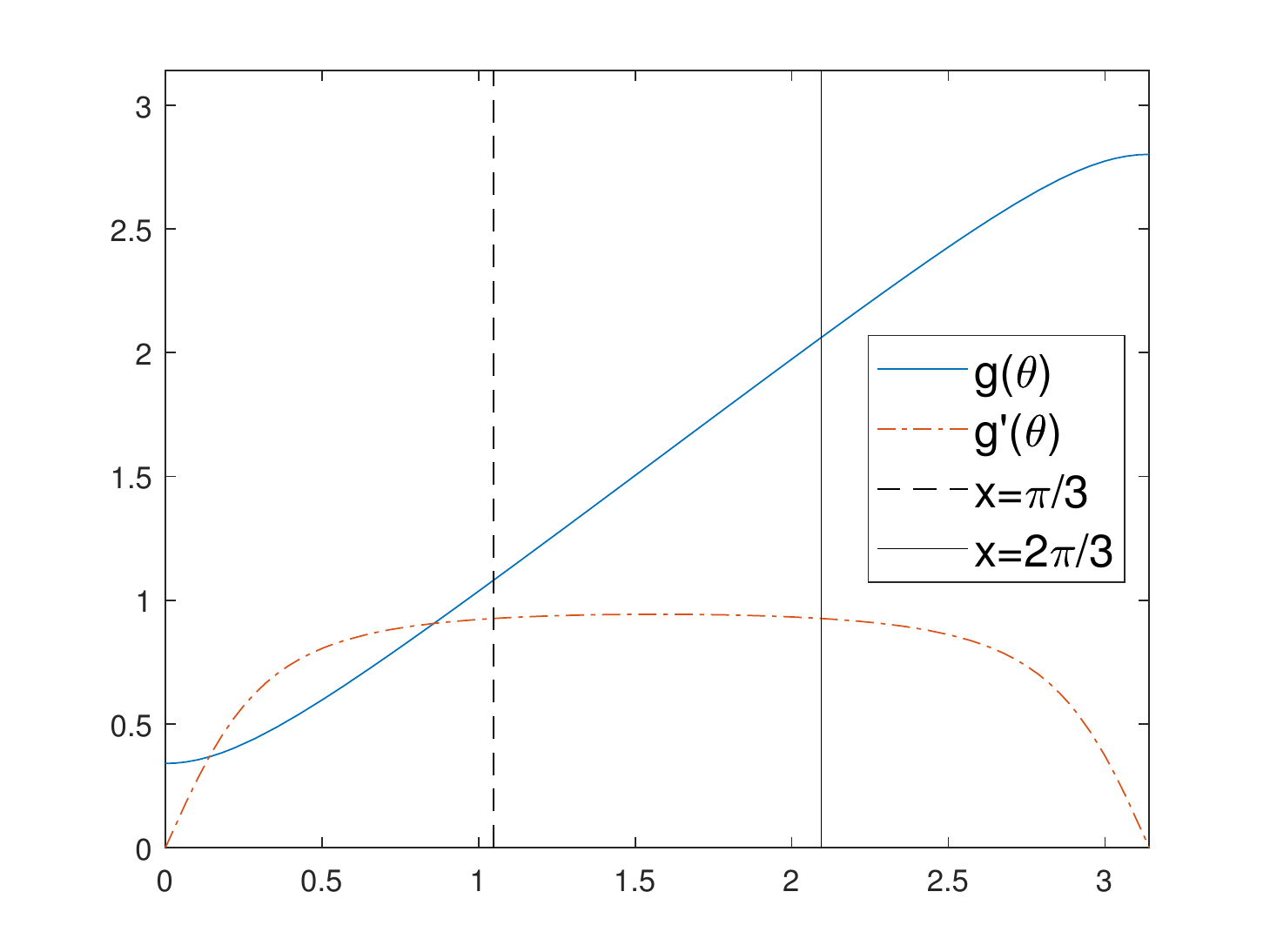}
	\caption{\label{fig:g_theta} The function $\varphi = g(\theta)$ defined in Eq.~\eqref{eq:g_def}, and its derivative $g'(\theta)$.}
\end{figure}

We now select the eigenvalue angle subset $S$ in Lemma~\ref{lem:irwin} to be those $\varphi_{\pm k}=g(\theta_{\pm k})$ whose $\theta_{\pm k} \in S' := (\frac{\pi}{3},\frac{2\pi}{3})$, where $g$ is the following function:
\begin{equation}\label{eq:g_def}
    g(\theta_{\pm k}) :=\arccos( 2\sqrt{pq} \cos\theta_{\pm k}).
\end{equation}
Note that $g(\theta) \in [\arccos(2\sqrt{pq}), \pi - \arccos(2\sqrt{pq})]$ when $\theta\in [0,\pi]$. 
Thus the angles $\varphi_{\pm k}$  with $ k=2\sim(n+1)$ have a constant gap with $\varphi_{\pm 1} = 0,\pi$.
Furthermore, it can be seen from Fig.~\ref{fig:g_theta} that $g(\theta)$ is monotone increasing on $(0,\pi)$, and the minimal derivative when $\theta\in S'$ is $g'(\pi/3)=\sqrt{6/7}$, since
\begin{equation}\label{eq:g_derivative}
    g'(\theta) = \frac{2\sqrt{pq} \sin\theta}{\sqrt{1-4pq\cos^2\theta}}.
\end{equation}
The minimal angle gap
\begin{equation}
    \Delta_\theta := \min \{ |\theta_{\pm k} - \theta_{\pm j}| : \theta_{\pm k} \in S', \theta_{\pm j} \in (0,\pi), k\neq j\}
\end{equation}
of $\theta_k$ will result in the minimal angle gap of $\varphi_k$ to satisfy:
\begin{equation}\label{eq:relation_theta_phi}
    \Delta E_S \geq \sqrt{6/7} \Delta_\theta,
\end{equation}
when $n$ is sufficiently large. This is because we only need to consider the gap between $\theta_{\pm k}$ and $\theta_{\pm j}$ all belonging to $S'$, and the gap $\overline{\Delta_\theta}$ between $\theta_{\pm k}\in S'$ nearest to $\pi/3$ and its adjacent $\theta_{\pm j}$ on the left.
In the latter case, we have $g(\theta_{\pm k}) - g(\theta_{\pm j}) \geq g'(\pi/3 - \overline{\Delta_\theta}) \overline{\Delta_\theta}$. But as $g'(\pi/3 - \overline{\Delta_\theta}) \to g'(\pi/3)$ when $n \to \infty$, Eq.~\eqref{eq:relation_theta_phi} holds when $n$ is sufficiently large.
Therefore, in order to bound $\Delta E_S$, it's sufficient to consider $\Delta_\theta$.

Eq.~\eqref{eq:eigval_eqn_2} in Lemma~\ref{lem:spectral_decomp} shows that the angles $\theta_{\pm k}$ where $ k=2\sim n+1$ are the $2n$ roots of the following equation in the interval $(0,\pi)$.
\begin{equation}\label{eq:eigval_eq_stoc}
    \frac{\sin(n+1)\theta}{\sin n\theta} = \mp \frac{1}{\sqrt{2}}.
\end{equation}

\begin{Remark}
    Equation~\eqref{eq:eigval_eq_stoc} is almost the same as the one presented in Ref.~\cite{CCD03}, but the RHS there is $\pm \sqrt{2}$.
    Therefore our analysis shown below is slightly different from those shown in Ref.~\cite{upper_improve} (which contains a review and some improvements of the results in Ref.~\cite{CCD03}).    
\end{Remark}

\begin{figure}[htbp]
	\centering
	\includegraphics[width=0.5\textwidth]{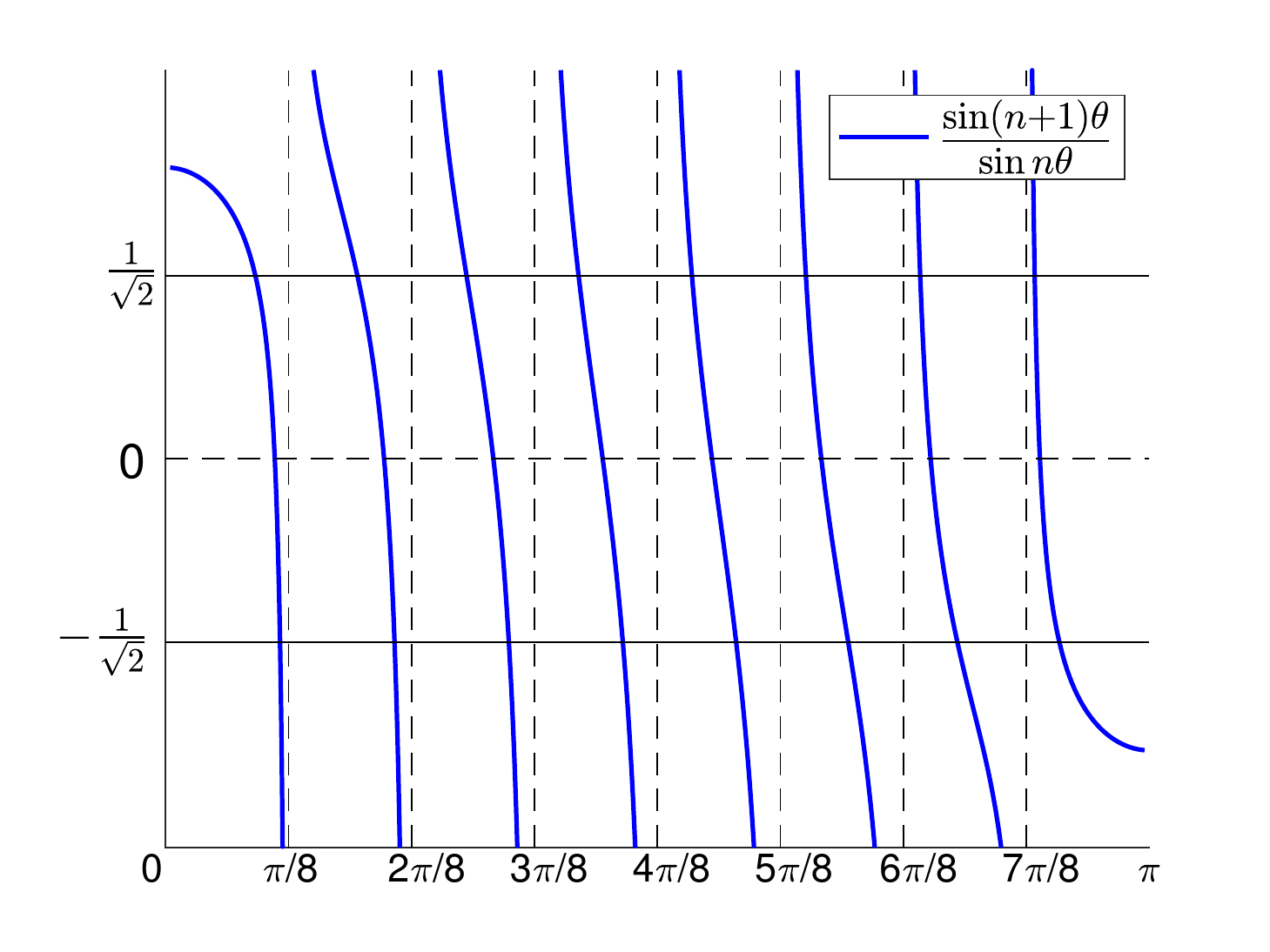}
	\caption{\label{fig:roots} Left hand side of Eq.~\eqref{eq:eigval_eq_stoc} for $n=8$.}
\end{figure}

For an intuitive understanding of Eq.~\eqref{eq:eigval_eq_stoc}, its LHS for $n=8$ and the horizontal lines $y=0,\pm \frac{1}{\sqrt{2}}$  are plotted in Fig.~\ref{fig:roots}.

We now present Lemma~\ref{lem:phase_gap} showing that $\Delta_\theta = \Omega(1/n)$ with our choice of $S'$.

\begin{Lemma}\label{lem:phase_gap}
    When $n$ is sufficiently large, the minimal gap $\Delta_\theta$ of the roots of Eq.~\eqref{eq:eigval_eq_stoc} has the following $\Omega(1/n)$ lower bound with $S' = (\frac{\pi}{3},\frac{2\pi}{3})$:
    \begin{align}
        \Delta_\theta & = \min \{ |\theta_{\pm k} - \theta_{\pm j}| : \theta_{\pm k} \in S', \theta_{\pm j} \in (0,\pi), k\neq j\} \label{eq:def_delts_theta}\\
        &\geq  \frac{\pi -2\theta_0}{n} \approx \frac{0.15\pi}{n},
    \end{align}
    where $\tan\theta_0 =\frac{\sqrt{3}}{\sqrt{2}-1}$.
\end{Lemma}

\begin{proof}
    By centrosymmetry, we only need to consider the angle gap between the root $\theta =\frac{l\pi}{n} -\delta$ corresponding to RHS$=-\frac{1}{\sqrt{2}}$ in Eq.~\eqref{eq:eigval_eq_stoc} and the root $\theta'=\frac{l'}{n}\pi+\delta'$ corresponding to RHS$=\frac{1}{\sqrt{2}}$ in Eq.~\eqref{eq:eigval_eq_stoc}, where $l' \in \{ l-1,l\}$.
    Since the $(n-1)$ zeros $\frac{l\pi}{n}$ of $\sin n\theta$ correspond to the vertical asymptotes of LHS of Eq.~\eqref{eq:eigval_eq_stoc}, we have $\delta,\delta' \in(0,\frac{\pi}{n})$.

    We now consider the lower and upper bound of $\delta$. Substituting $\theta =\frac{l\pi}{n} -\delta$ into Eq.~\eqref{eq:eigval_eq_stoc}, and using the trigonometric identity $\sin(a-b) =\sin(a) \cos(b) -\cos(a) \sin(b)$, we have
    \begin{align}
        -\sqrt{2} \sin(n\theta +\theta) &= \sin(n\theta) \\
        \Leftrightarrow -\sqrt{2} \sin(l\pi -n\delta +\frac{l\pi}{n} -\delta) &= \sin(l\pi -n\delta)\\
        \Leftrightarrow -\sqrt{2} \sin(n\delta - \frac{l\pi}{n} +\delta) &= \sin(n\delta)\\
        \Leftrightarrow -\sqrt{2} \sin(n\delta - \theta) &= \sin(n\delta)\\
        \Leftrightarrow -\sqrt{2} [\sin(n\delta) \cos\theta -\cos(n\delta) \sin\theta] &= \sin(n\delta)\\
        \Leftrightarrow -\sqrt{2} [\tan(n\delta) \cos\theta -\sin\theta] &= \tan(n\delta)
    \end{align}
    Thus
    \begin{equation}\label{eq:root_eq_minus}
        \tan(n\delta) = \frac{\sqrt{2} \sin\theta}{1 +\sqrt{2}\cos\theta}.
    \end{equation}
    Since the RHS of Eq.~\eqref{eq:root_eq_minus} is monotone increasing on $(\frac{\pi}{3} ,\frac{2\pi}{3} ) \ni \theta$, we have $\frac{\sqrt{3}}{\sqrt{2}+1} < \tan(n\delta) < \frac{\sqrt{3}}{\sqrt{2}-1}$, from which $\frac{\theta_1}{n} < \delta < \frac{\theta_0}{n}$, where $\theta_1 =\arctan\frac{\sqrt{3}}{\sqrt{2}+1}$.

    We now consider the lower and upper bound of $\delta'$. If we Substitute $\theta'=\frac{l'}{n}\pi+\delta'$ into Eq.~\eqref{eq:eigval_eq_stoc}, we have
    \begin{equation}
        \tan(n\delta') = \frac{\sqrt{2} \sin\theta'}{1 -\sqrt{2}\cos\theta'}.
    \end{equation}
    Similarly, it can be shown that $\frac{\theta_1}{n} < \delta' < \frac{\theta_0}{n}$. 
    Therefore, $\Delta_\theta \geq \min\{ \frac{\pi}{n} - 2\frac{\theta_0}{n}, 2\frac{\theta_1}{n}\} =\frac{\pi-2\theta_0}{n} \approx \frac{0.15\pi}{n}$.
\end{proof}

\subsection{Proof of Theorem~\ref{thm:main}}\label{subsec:proof_main_thm}

With the help of the above three Lemmas~\ref{lem:spectral_decomp}-\ref{lem:phase_gap}, We can now prove Theorem~\ref{thm:main}.

    In order to lower bound the items $|y_j^* c_j|$ of the sum in Eq.~\eqref{eq:aver_pr_lower} in Lemma~\ref{lem:irwin}, we first calculate:
    \begin{align}
        &\braket{2n+1,L | u_{\pm,\pm k}} \braket{u_{\pm,\pm k} | 0,R} \\ 
        &= \bra{2n+1,L} \cdot (\ket{a_{\pm k}} - \exp(\pm i\varphi_{\pm k}) \ket{b_{\pm k}}) \cdot 
        (\bra{a_{\pm k}} - \exp(\pm i\varphi_{\pm k}) \bra{b_{\pm k}}) \cdot \ket{0,R} \\
        &= [v_{\pm k}(1) -e^{\pm i\varphi_{\pm k}} \sqrt{p} v_{\pm k}(2)] \cdot [v_{\pm k}(1) -e^{\mp i\varphi_{\pm k}} \sqrt{p} v_{\pm k}(2)] \cdot(\pm_k) \\
        &= [v_{\pm k}(1)^2 +p v_{\pm k}(2)^2 -2\lambda_{\pm k} v_{\pm k}(1) v_{\pm k}(2)] \cdot(\pm_k) \\
        &= (\pm_k)(1-\lambda_{k}^2). \label{eq:product_result}
    \end{align}
    The last line follows from $v_{\pm k}(1)=1$ and $ v_{\pm k}(2)=\lambda_{\pm k}/\sqrt{p}$.
    Therefore, combined with the norm $\|\ket{u_k}\|$ shown by Eq.~\eqref{eq:v_k_norm} in Lemma~\ref{lem:spectral_decomp}, we have the following identities when $\theta_k \in S'$.
    \begin{align}
        |\bra{2n+1,L} \Pi_{\ket{u_{\pm,\pm k}}} \ket{0,R}| &= \frac{1-\lambda_{k}^2}{ \|\ket{u_k} \|^2} \\
        &= \frac{q}{2} \cdot \frac{\sin^2\theta_k}{1-\lambda_k^2} \Big( n +\sqrt{\frac{q}{p}} \frac{\sin((n+1)2\theta_k)}{2\sin\theta_k} \Big)^{-1} \\
        &\geq \frac{1}{3} \cdot \frac{27}{28} \cdot (n + \sqrt{2} \frac{1}{2\cdot\sqrt{3}/2})^{-1} \\
        &\geq \frac{9}{28n} + O(\frac{1}{n^2}).
    \end{align}
    The third line above follows from $\theta_k \in (\frac{\pi}{3},\frac{2\pi}{3})$.
    Since $\theta_k$ is almost uniformly distributed in $(0,\pi)$,
    we have $\sum_{j\in S} |y_j^*\, c_j|^2 \geq \frac{2}{3}\cdot 4n \cdot (\frac{9}{28n})^2 > \frac{1}{4n}$.
    
    We set 
    \begin{equation}
        k \geq \log 5n,
    \end{equation}
    and
    \begin{equation}
        T \geq \frac{\pi}{2 \Delta E_S},
    \end{equation}
    which is approximately $\frac{n}{0.3\sqrt{6/7}} \approx 3.6 n$ by Lemma~\ref{lem:phase_gap} and Eq.~\eqref{eq:relation_theta_phi}. Then $\left( \frac{\pi}{T\, \Delta E_S} \right)^k \leq (\frac{1}{2})^{\log 5n} = \frac{1}{5n}$.    
    And thus by Eq.~\eqref{eq:aver_pr_lower},
    \begin{align}
        p(y|\psi_0) &\geq \sum_{j\in S} |y_j^*\, c_j|^2 -\Big( \frac{\pi}{T\, \Delta E_S} \Big)^k \\ 
        &\geq \frac{1}{4n} - \frac{1}{5n} = \frac{1}{20n}.
    \end{align}
    Note that $p(t) = \left|\bra{4n+1} M_U^t \ket{0}\right|^2 = 0$ when $t<2n$.
    In fact, $M_U \ket{0} = \ket{1}$ and one iteration of $M_U$ can propagate the amplitude from $\ket{k}$ to at furthest $\ket{k+2}$, thus $p(t) = 0$ when $t < 2n$.
    Since maximum is greater than average, we have now proven Theorem~\ref{thm:main}.

\begin{Remark}\label{rem:fixed_t}
We can actually obtain an explicit expression of $A(t) := \bra{4n+1} M_U^t \ket{0}$ for odd $t$ as
\begin{align}\label{eq:A_t}
A(t) 
&= \frac{p-q}{2p-(q/p)^{n}} +2q\sum_{k=2}^{n+1}\frac{ \cos(t\arccos(2\sqrt{pq}\cos\theta_k))} {1-4pq\cos^2\theta_k } 
\frac{\sin^2\theta_k}{ n +\sqrt{\frac{q}{p}} \frac{\sin((n+1)2\theta_k)}{2\sin\theta_k} }.
\end{align}
Since the expression is too complicated, our first attempt of directly analyzing $A(t)$ fails, and thus we have turned to the help of Lemma~\ref{lem:irwin}.
\end{Remark}
$A(t)$ can be calculated as follows.
By Eq.~\eqref{eq:product_result} and the expression of $\ket{u_{\pm 1}}$ shown in Lemma~\ref{lem:spectral_decomp}, we have
\begin{align}
    A(t) &= \sum_{\pm} \sum_{\pm_k} \sum_{k=2}^{n+1} e^{\pm it\varphi_{\pm k}}
    \frac{(\pm_k)(1-\lambda_k^2)}{\| \ket{u_k} \|^2}
    + \sum_{\pm} (\pm)^t 
    \frac{(\pm)}{\|\ket{u_{1}}\|^2}.
\end{align}
Note that
\begin{equation}
    \sum_{\pm} e^{\pm i t \varphi_{\pm k}} = 2\cos(t\varphi_{\pm k}).
\end{equation}
Since $\varphi_{-k} = \pi - \varphi_k$, we have
\begin{equation}
    \sum_{\pm_k} \cos(t \varphi_{\pm k}) (\pm_k) = \cos(t\varphi_k) (1-(-1)^t).
\end{equation}
Similarly,
\begin{equation}
    \sum_{\pm} (\pm)^t (\pm) =1-(-1)^t.
\end{equation}
Therefore, it can be seen that
\begin{equation}\label{eq:even_zero}
    A(t)=0, \quad \text{if} \  t \ \text{is even.}
\end{equation}
Thus we will only consider odd $t$.
Substituting the square of norms $\|\ket{u_{\pm k}}\|^2,k=1\sim (n+1)$ in Lemma~\ref{lem:spectral_decomp}, we have $A(t)$ as shown in Eq.~\eqref{eq:A_t}.

\section{Deterministic quantum algorithm   with $O(n^{1.5}\log n)$ queries}\label{sec:alg}
With the matrix expression $M_U = M_S \cdot M_C$ of $U_\mathrm{walk}$ within its reduced invariant subspace $\mathcal{H}_0$  obtained in Section~\ref{sec:subspace}, the exact value of the amplitude on the target state after $T_1$ steps of quantum walks, i.e. $\bra{4n+1} M_U^{T_1} \ket{0} $, can be calculated exactly.
Therefore, combining with one of the deterministic quantum search algorithms \cite{exact}, for example Long's algorithm \cite{Long}, we can design a deterministic quantum algorithm for the welded tree problem as shown in Algorithm~\ref{alg:main}.

More precisely, if there is a quantum process (unitary operation) $\mathcal{A}$ that transform some initial state $\ket{0}$ to $\ket{\psi}$ that has known overlap, i.e. $p = |\braket{t | \psi}| \in (0,1)$, with the desired target state $\ket{t}$.
Then Long's algorithm \cite{Long} can amplify the overlap to $1$ by applying the generalized Grover's iteration $G(\alpha,\beta)=\mathcal{A} S_0(\beta) \mathcal{A}^\dagger \cdot S_t(\alpha)$, where $S_t(\alpha)=e^{i\alpha \ket{t}\bra{t}}$ and $S_0(\beta)=e^{-i\beta \ket{0}\bra{0}}$, for $T = O(1/p)$ times to the state $\ket{\psi}=\mathcal{A}\ket{0}$.

In the case of the original Grover's algorithm \cite{Grover}, $\mathcal{A} := H^{\otimes n}$ is the Hadarmard gates on $n$ qubits, $\ket{0} := \ket{0}^{\otimes n}$, and $\ket{t}$ is the equal-superposition of all target elements.
In the case of Algorithm~\ref{alg:main}, $\mathcal{A} := U_\mathrm{walk}^{T_1} U_p$, $\ket{0} := \ket{s,\perp}$, and $\ket{t} := \ket{t,\varphi(t)}$.
The parameters $T,\alpha,\beta$ are determined by the known overlap $p$, where in Grover's case is the square root of the proportion of target elements in the unstructured database, and in our case is $p_{T_1} = |\bra{4n+1} M_U^{T_1} \ket{0}| $.

\begin{algorithm}[htp]
	\caption{Deterministic quantum algorithm for the welded tree problem}
	\label{alg:main}
	\begin{description}
		\item[Input:] adjacency list quantum oracle $O$ (see Eq.~\eqref{eq:oracle}) for the welded tree $G_n$, and the entrance name $s \equiv 0^{2n}$.
		\item[Output:] the exit name $t$.
		\item[Procedure:]
	\end{description}
	\begin{enumerate}
		\item Same as step 1 in Algorithm~\ref{alg:succinct}.
		\item Construct quantum circuit of the generalized Grover's iteration $G(\alpha,\beta)= \mathcal{A} S_0(\beta) \mathcal{A}^\dagger \cdot S_t(\alpha)$ as in Fig.~\ref{fig:grover}, where $\mathcal{A} := U_\mathrm{walk}^{T_1} U_p$.
        Set the parameters $\alpha, \beta, T_2$ as $\alpha = -\beta = 2\arcsin \left( \frac{\sin\frac{\pi}{4T_2+2} }{\sin\theta } \right)$ and $T_2 = \lceil \frac{\pi/2-\theta}{2\theta} \rceil$, where $\theta = \arcsin(p_{T_1})$.
		\item  
		Apply $G(\alpha,\beta)^{T_2}$ to $\ket{\psi_{T_1}} = \mathcal{A} \ket{s,\perp}$. This will result in $\ket{t,\varphi(t)}$ exactly and thus measuring register $r_1$ leads to the exit name $t$ with certainty.
	\end{enumerate}
\end{algorithm}

\begin{figure}
	\centering
	\includegraphics[width=0.7\textwidth]{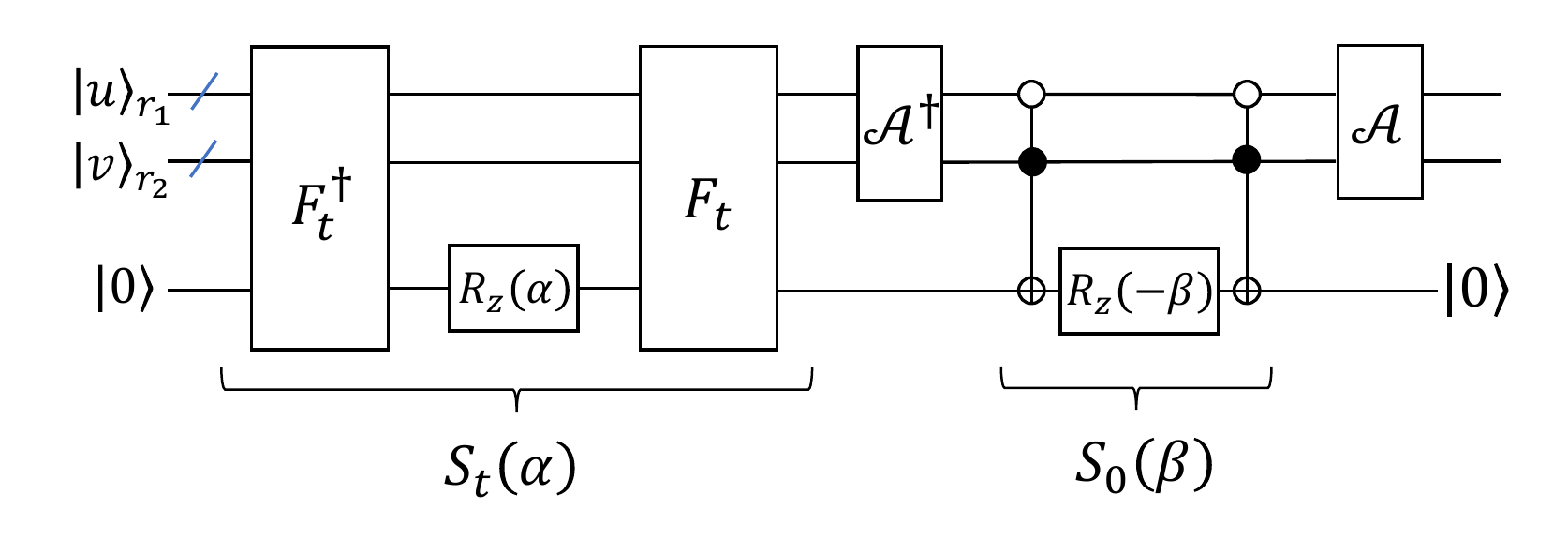}
	\caption{\label{fig:grover} Quantum circuit implementing the generalized Grover's iteration $G(\alpha,\beta)= \mathcal{A} S_0(\beta) \mathcal{A}^\dagger \cdot S_t(\alpha)$. Single qubit operation $R_z(\alpha) := \mathrm{diag}(1,e^{i\alpha})$. Operator $S_0(\beta)$ add phase shift $e^{-i\beta}$ to the initial state $\ket{s,\perp} = \ket{0^{2n},1^{2n}}$, and operator $F_t$ flip the last auxiliary qubit conditioned on $\ket{u}_{r_1} =\ket{t}$. Determining that $u=t$ can be done by querying the three adjacent vertices of $u$ and checking that exactly one of them is $\perp$ and $u\neq s$. }
\end{figure}

\section{Numerical simulation supporting $O(n^{4/3})$ query complexity }\label{subsec:numeri}

We find that the actual performance of our algorithms is better than $O(n^{1/2} \cdot n \log n)$, since results of numerical simulation (conducted by MATLAB) show that the success amplitude will be $\Omega(n^{-1/3})$ when $U_\mathrm{walk}$ is applied  for $O(n)$ times.
We formalize it in the following Conjecture~\ref{conj:walk_prob} (an improved version of Theorem \ref{thm:main}). Then the query complexity of Algorithm~\ref{alg:main} will be $O(n^{1/3} \cdot n) = O(n^{4/3})$.

\begin{Conjecture}\label{conj:walk_prob}
There exists odd number $T \in [ 2n, 2.5n ]$ ($T\approx n/\sqrt{pq} \approx 2.1213n$ for sufficient large $n$) such that
	\begin{equation}\label{eq:p_T}
		p_{T} := \left|\bra{4n+1}  M_U^{T} \ket{0} \right| > n^{-1/3}.
	\end{equation}
\end{Conjecture}
\begin{Remark}
    We only consider \textit{odd} $T$ because Eq.~\eqref{eq:even_zero} shows that $p_T = 0$ when $T$ is even.
    The constant $\sqrt{pq}$ may come from the fact that $g'(\pi/2)= 2\sqrt{pq}$ (see Eq.~\eqref{eq:g_derivative}) and $2\pi/g'(\pi/2)=1/\sqrt{pq}$.
\end{Remark}

One can obtain the exact value of the largest $p_T$ with $T\in [2n,2.5n]$ (denoted by $P_T$) using, for example, MATLAB's Symbolic Math Toolbox. The exact value of $P_T$ when $n=50,100,150$ is shown in Table~\ref{tab:numerical}.

\begin{table}[ht]
	\centering
	\begin{tabular}{lll}
		\toprule
		$n$ & $T$ & $P_T$ \\
		\midrule
		$50$
		& $109$ 
		& $2^{152} \cdot 19 \cdot 38861 / 3^{108}$ \\
		$100$
		& $215$
		& $2^{300} \cdot 318388779301 / 3^{214}$ \\
		$150$
		& $323$
		& $2^{451} \cdot 274739 \cdot 1231103390273 / 3^{322}$ \\
		\bottomrule
	\end{tabular}    
	\caption{The exact value of $P_T$ when $n=50,100,150$.}
	\label{tab:numerical}
\end{table}

\begin{figure}[ht]
	\centering
    \includegraphics[width=0.5\textwidth]{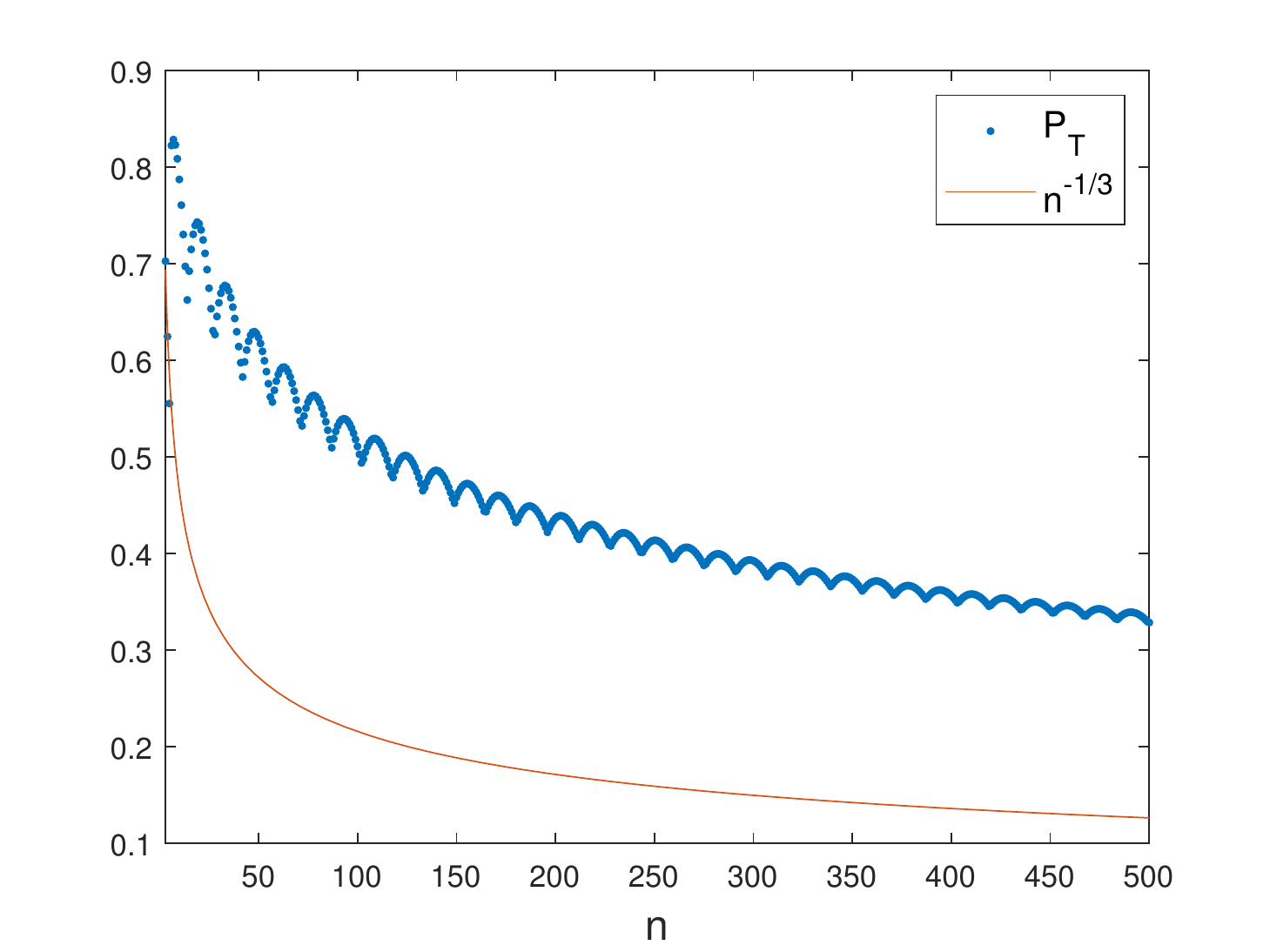}
	\caption{\label{fig:fig 1} Scatter diagram of $P_T$ for $n=3,\cdots,500$. Orange solid line represents $n^{-1/3}$.}	
\end{figure}

The scatter diagram of $P_T$ for $n=3,\cdots,500$ is shown in Fig.~\ref{fig:fig 1}. It can be seen that $P_T>n^{-1/3}$ holds for all $n\in [6,500]$, which supports Conjecture~\ref{conj:walk_prob}.

\begin{figure}[ht]
	\centering
	\includegraphics[width=0.5\textwidth]{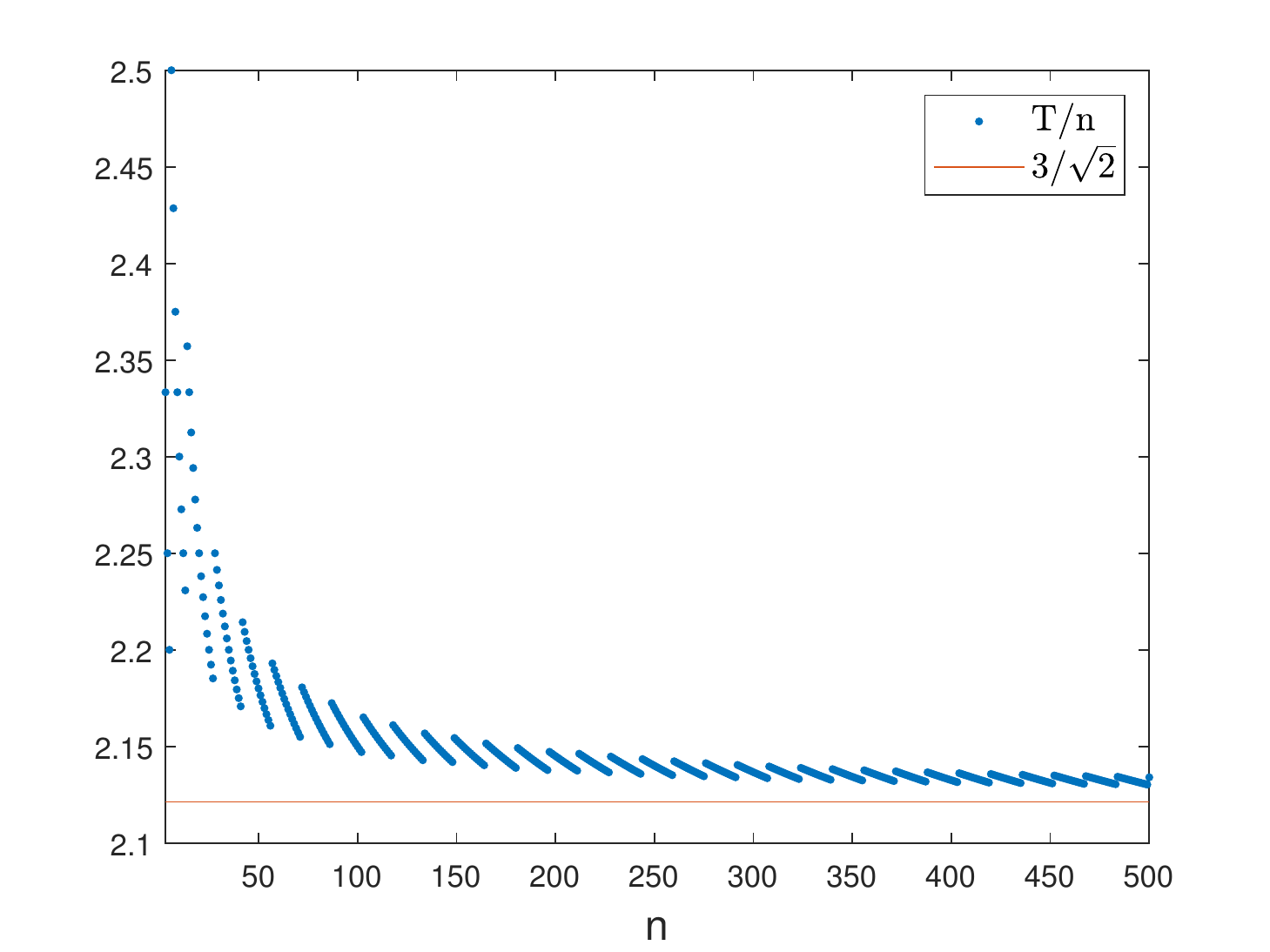}
	\caption{\label{fig:fig 2} Scatter diagram of $T/n$ for $n=3,\cdots,500$. The horizontal line represents $\frac{1}{\sqrt{pq}} =\frac{3}{\sqrt{2}}$.}
\end{figure}

The scatter diagram of the ratio $T/n$ for $n=3,\cdots,500$ is shown in Fig.~\ref{fig:fig 2}.
It can be seen that $T/n$ tends to $\frac{1}{\sqrt{pq}} =\frac{3}{\sqrt{2}} \approx 2.12$ as $n\to \infty$.

\begin{figure}[ht]
	\centering
	\includegraphics[width=0.5\textwidth]{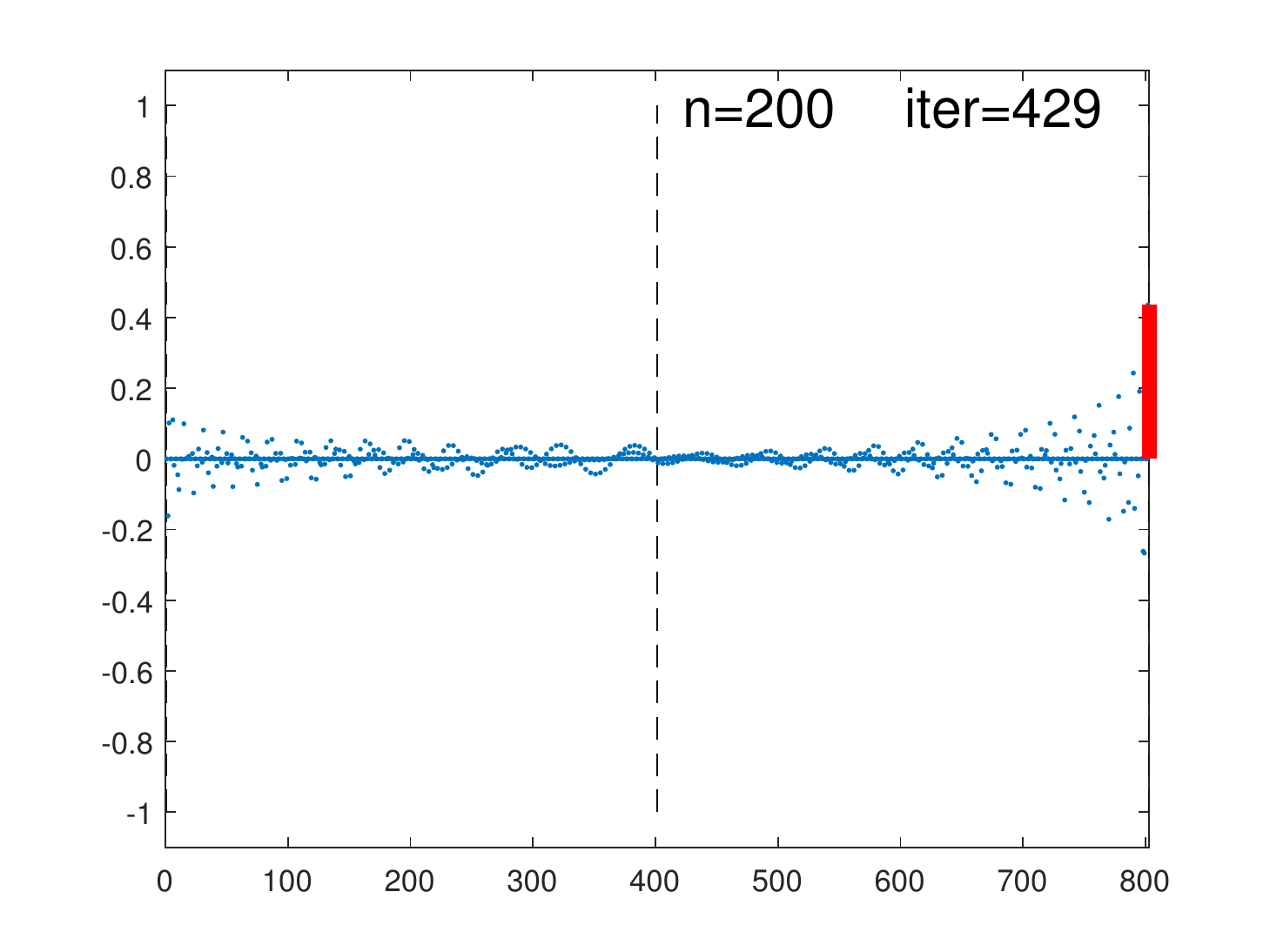}
	\caption{\label{fig:video} The frame when $T=429$ of the video \cite{video} showing the evolution of the $(4n+2)$-dimensional vector $\ket{\psi_T}$ with $n=200$. The $x$-axis represents the indices $k\in\{0,\cdots,4n+1\}$ of $\ket{\psi_T}$, and the $y$-axis represents the amplitude of the $k$-th component, which is between $[-1,1]$. The red pillar on the rightmost represents the amplitude $\braket{4n+1 \vert \psi_{T}}$.}
\end{figure}

We also depict, as an example $n=200$, evolution of the $(4n+2)$-dimensional vector $\ket{\psi_T} = M_U^{T} \ket{0}$ for odd $T$ ranging from $1$ to $2.5n$ in video~\cite{video}.
The frame when $T=429$ of the video is shown in Fig.~\ref{fig:video}.
It can be seen that the amplitude of the state vector $\ket{\psi_T}$ can be positive or negative, showing the periodic coherent and destructive nature of quantum walk.

\section{Conclusion}\label{sec:conclu}
In this paper, we have revisited quantum algorithms for the welded tree problem and proposed a rather succinct algorithm based purely on the simplest coined quantum walks.
A rigorous polynomial query upper bound is provided based on spectral decomposition of the reduced quantum walk matrix.
The succinctness of our algorithm re-displays the power of the simplest framework of quantum walks, changing the stereotype that the existing DTQW frameworks before the multidimensional one can achieve at most a quadratic speedup over the best classical algorithm.
Our algorithm for the welded tree problem can also be made zero-error theoretically, making it one of the few examples of an exponential separation between the zero-error (exact) quantum and the randomized query complexities.
Numerical simulation indicates that the actual performance of our algorithms is better, and it will be a future work to prove Conjecture~\ref{conj:walk_prob}.

\appendix
\section{Implementing $U_\varphi$ in Lemma~\ref{lem:ref_A}}\label{sec:coin_imp}
In this appendix, we will implement the unitary operator $U_\varphi: \ket{u,\perp} \mapsto \ket{u, \varphi(u)}$ in Lemma~\ref{lem:ref_A}.
We first introduce five auxiliary registers 
    \begin{equation}
        \ket{0}_{q_1} \ket{0}_{q_2} \ket{0}_{q_3} \ket{0}_a \ket{0}_b \ket{0}_c,
    \end{equation}
    where register $q_i$ consists of $2n$ qubits storing the query result, register $a$ is a qutrit with state space $\mathbb{H}^3=\text{span} \{\ket{0},\ket{1},\ket{2} \}$ used for generating $\ket{\varphi(u)}$ when $u$ is an internal node, register $b$ is a qudit with state space $\mathbb{H}^5=\text{span} \{\ket{0},\cdots,\ket{4} \}$ storing conditions, and register $c$ is a qubit used for generating $\ket{\varphi(u)}$ when $u$ is one of the roots.
	Now $U_\varphi$ can be implemented as follows, where oracle query happens in the first and last step.
	\begin{enumerate}[1.]
        \item Query the oracle $O$ on registers $r_1,q_1,q_2,q_3$ to obtain
		\begin{equation}\label{eq:U_A_begin}
			\ket{u}_{r_1} \ket{\perp}_{r_2} \bigotimes_{i=1}^{3} \ket{\Gamma(u,i)}_{q_i}\ \ket{0}_a \ket{0}_b \ket{0}_c.
		\end{equation}

        \item Apply the transformation $\ket{q}\ket{b} \mapsto \ket{q} \ket{b+f(q)}$ on register $q:=(q_1,q_2,q_3)$ and $b$. The function $f: \{0,1\}^{6n} \to \{0,\cdots,4\}$ is defined as: $f(q)=0$ iff there's no $q_i=\perp$, so that $u$ is an internal node; $f(q)=i$ for $i=1,2,3$ iff there's one and only one $q_i=\perp$, so that $u \in \{s,t\}$, and the $i$th register $q_i$ stores the value $\perp$; $f(q)=4$ iff there's more than one $q_i=\perp$, so that $u\notin V(G_n)$.
        \textit{It can be easily seen that calculating $f$ takes $O(n)$ basic operations.}

        \item Conditioned on $b=0$, i.e. $u$ is an internal node, apply the following two steps.

        \begin{enumerate}
        \item[3.1.] Flip all the qubits of register $r_2$ so that it's set to $\ket{0^{2n}}$. Apply quantum Fourier transform $QFT_3$ to register $a$, and then controlled by $\ket{i}_a,\, i\in\{0,1,2\}$, add (i.e. bit-wise modulo $2$ addition) the value of register $q_{(i+1)}$ to register $r_2$, obtaining
		\begin{equation}
			\ket{u}_{r_1} (\frac{1}{\sqrt{3}}\sum_{i=0}^2 \ket{\Gamma(u,i+1)}_{r_2} \ket{i}_a)\ \bigotimes_{i=0}^{2} \ket{\Gamma(u,i+1)}_{q_{(i+1)}} \ket{b}_b \ket{0}_c.
		\end{equation}
		\textit{This controlled addition can be done in $O(n)$ basic operations.}
        \item[3.2.] Compare $\ket{\Gamma(u,i)}_{r_2}$ with $\ket{\Gamma(u,j)}_{q_j}$ for $j=1,2,3$ and subtract $\ket{i}_a$ with $(j-1)$, where $j$ is the \textit{unique} index $j$ such that $\ket{\Gamma(u,i)}_{r_2}=\ket{\Gamma(u,j)}_{q_j}$, obtaining
		\begin{equation}\label{eq:U_A_end}
			\ket{u}_{r_1} \ket{\varphi(u)}_{r_2} \bigotimes_{i=1}^{3} \ket{\Gamma(u,i)}_{q_{i}}\ \ket{0}_a \ket{b}_b \ket{0}_c.
		\end{equation}
        The uniqueness of the index $j$ can be easily seen from the condition that all of $u$'s neighbours $\Gamma(u,i)$ are distinct.
        \textit{This compare (between binary strings) and subtract operation can be done in $O(n)$ basic operations.}
        \end{enumerate}

        \item Conditioned on $b\in \{1,2,3\}$, i.e. $u$ is one of the two roots and register $q_b$ stores $\perp$, apply the following steps.
    
        \begin{enumerate}
            \item[4.1.] Swap register $q_b$ and $q_3$ so that the first two auxiliary registers store the genuine adjacent vertex name of $u \in \{s,t\}$.
            \textit{The conditioned SWAP operation can be done in $O(n)$ basic operations.}
            
            \item[4.2.] similar to step 3.1 and 3.2, transform $\ket{\perp}_{r_2}$ to $\ket{\varphi(u)}_{r_2}$ with the help of $H \ket{0}_c = \frac{1}{\sqrt{2}}(\ket{0}+\ket{1})$.
            
            \item[4.3.] Repeat step 4.1 so that the order of register $q_i$ is restored, ensuring the success of step~7.
        \end{enumerate}
        
        \item Conditioned on $b = 4$, apply the identity transformation $I$, since register $r_1$ already stores $\varphi(u)=\perp$ (the three `neighbours' of $u\notin V(G_n)$ are all $\perp$).
        
        \item Similar to step 2, apply the transformation $\ket{q}\ket{b} \mapsto \ket{q}\ket{b-f(q)}$ to register $q,b$, where the subtraction is modulo $5$. Therefore, register $b$ is recovered to $\ket{0}_b$.
        
		\item Query the oracle $O$ once more as in step 2, so that all the auxiliary registers are restored back to zero.
	\end{enumerate}
Thus $U_\varphi$ can be implemented with $2$ oracle queries and \textit{$O(n)$ basic operations}.

\section{Spectral decomposition of the reduced matrix}\label{sec:spectral}
In this Appendix, we prove Lemma~\ref{lem:spectral_decomp} about the spectral decomposition of the reduced coined quantum walk matrix $M_U = M_S M_C$. The proof is inspired by~\cite{localized}.
However, since its analysis is not perfect (as they did not obtain analytical expression of $U_k$ shown in Eq.~\eqref{eq:U_k_cheby} by comparing with Chebyshev polynomial of the second kind), we will show in detail the complete proof for the sake of completeness and convenience of the readers.

We also expand or improve some of the implicit or complicated steps in \cite{localized}, and point out a connection with another commonly used technique for analyzing quantum walk operators, i.e. the singular value decomposition, or more precisely, Jordan's Lemma \cite{Jordan_1875} about common invariant subspaces of two reflection operator, which has been used in Refs.~\cite{Szegedy_03, KroviMOR16, QSVT, eedp}.

We first present the following helper Lemma~\ref{lem:U_convert}, which is implicit in~\cite{localized} and similar to~\cite[Theorem 1]{childs2010relationship}, saying that in order to obtain the spectral decomposition of $M_U = M_S M_C$, we can instead consider the spectral decomposition of the following matrix
\begin{equation}
    A^\dagger M_S A =: J_{2n}.
\end{equation}

\begin{Lemma}\label{lem:U_convert}
    Consider the quantum walk operator $U=\mathrm{Ref}_B\, \mathrm{Ref}_A$, where $\mathrm{Ref}_A = (2A A^\dagger -I)$ and $A$ is a matrix with full column rank satisfying $A^\dagger A =I$. Let
    \begin{equation}
        \ket{a} :=A\ket{v},\ \ket{b} := \mathrm{Ref}_B \ket{a},
    \end{equation}
    where $\|\ket{v}\|^2=1$. If
    \begin{equation}\label{eq:J_eig}
        A^\dagger \mathrm{Ref}_B A \ket{v} = \lambda \ket{v}.
    \end{equation}
    Then, when $\left| \lambda \right| < 1$, we have
    \begin{equation}
        U \ket{u} := U \big(\ket{a} -e^{\pm i\varphi} \ket{b} \big) = e^{\pm i\varphi} \ket{u},
    \end{equation}
    where $\varphi := \arccos\lambda$, and $\| \ket{u}\|^2 =2(1-\lambda^2)$.
    And when $\lambda = \pm 1$, we have
    \begin{equation}
        U \ket{a} = \pm \ket{a}.
    \end{equation}
\end{Lemma}

\begin{proof}
    We first consider the case when $ \left| \lambda \right| < 1 $. From Eq.~\eqref{eq:J_eig} we know
    \begin{align}
        U \ket{b} &= \mathrm{Ref}_B (2AA^\dagger -I) (\mathrm{Ref}_B A \ket{v}) \\
        &= 2\mathrm{Ref}_B A \lambda \ket{v} -A\ket{v} \\
        &= 2\lambda \ket{b} -\ket{a}.
    \end{align}
    Therefore, $\mathrm{span} \{\ket{a},\ket{b}\}$ is an invariant subspace of $U$, and $U$ takes the following matrix form:
    \begin{equation}
        L=\begin{bmatrix}
            0 & -1 \\
            1 & 2\lambda
        \end{bmatrix}.
    \end{equation}
    Let $\lambda=\cos\varphi$, then we obtain the eigenvalues and eigenvectors of $L$: $e^{\pm i\varphi}$ and $[1,-e^{\pm i\varphi}]^T$. This can be easily verified by the following identities:
    \begin{align}
        1-2\lambda e^{\pm i\varphi}
        &= 1 -2\cos(\pm \varphi) e^{\pm i\varphi} \\
        &= 1 -2\cos^2(\pm\varphi) -2i\cos(\pm\varphi) \sin(\pm\varphi) \\
        &= -\cos(\pm 2\varphi) -i\sin(\pm 2\varphi) \\
        &= - e^{\pm 2i\varphi}
        = e^{\pm i\varphi}\cdot (-e^{\pm i\varphi}).
    \end{align}
    Therefore, we obtain two eigenvalues $e^{\pm i\varphi}$ of $U$ and their respective eigenvectors $\ket{u} :=\ket{a} -e^{\pm i\varphi} \ket{b}$.
    We now calculate the square of its norm:
    \begin{align}
        \braket{u|u} &=2-2\, \text{Re} ( e^{\pm i\varphi} \braket{a|b} )\\
        &= 2(1-\lambda^2).
    \end{align}
    The second line follows from $\braket{a|b}=\bra{v} A^\dagger \text{Ref}_B A \ket{v} =\lambda$ and the fact that $A^\dagger \text{Ref}_B A$ is Hermitian whose eigenvalue $\lambda$ is a real number.

    We now consider the case when $\lambda = \pm 1$. From Eq.~\eqref{eq:J_eig}, we know $\bra{v} A^\dagger \mathrm{Ref}_B A \ket{v} = \pm 1$, and thus $\mathrm{Ref}_B A \ket{v} = \pm A \ket{v}$. Therefore,
    \begin{align}
        U \ket{a} &= U A \ket{v} \\
        &= \mathrm{Ref}_B (2AA^\dagger - I) A \ket{v} \\
        &= \mathrm{Ref}_B A \ket{v} \\
        &= \pm A \ket{v} \\
        &=\pm \ket{a}.
    \end{align}
\end{proof}

\begin{Remark}
Suppose $\mathrm{Ref}_B = 2BB^\dagger -I$. Let $D :=A^\dagger B$, and consider its singular value decomposition $\sum_i s_i \ket{v_i} \bra{w_i}$, which is a common approach in Refs.~\cite{Szegedy_03, KroviMOR16, QSVT, eedp}. Then the connection between the eigenvalue $\lambda$ of $A^\dagger\, \mathrm{Ref}_B\, A$ and the singular value $s$ of $D$ is:
\begin{equation}\label{eq:connection}
    \arccos(\lambda) =2 \arccos(s).
\end{equation}
\end{Remark}

\begin{proof}
Since $DD^\dagger \ket{v_i} =s_i^2 \ket{v_i}$, we have
\begin{align}
A^\dagger\, \text{Ref}_B\, A &= A^\dagger (2BB^\dagger-I) A \\
&= 2(A^\dagger B) (A^\dagger B)^\dagger -I \\
&= 2DD^\dagger - I.
\end{align}
Therefore,
\begin{equation}
    A^\dagger\, \text{Ref}_B\, A \ket{v} =\lambda \ket{v} \Leftrightarrow DD^\dagger \ket{v} = \frac{\lambda+1}{2} \ket{v}.
\end{equation}
From the identity $ \frac{\cos\phi_i +1}{2} = \cos^2\frac{\phi_i}{2} $, we obtain Eq.\eqref{eq:connection}. 
\end{proof}

We now analyze the eigenvalues and eigenvectors of $J_{2n}=A^\dagger M_S A$.
First, we will need the matrix expression of $J_{2n}$.
Recalling $A$ as defined in Eq.~\eqref{eq:matrix_A}, and since $M_S$ swaps two adjacent rows (columns), we have:
\begin{equation}
    A^\dagger M_S=
\left[\begin{array}{ccccccc|c}
0 & 1 & & & & & & \\
\sqrt{p} & 0 & 0 & \sqrt{q} & & & & \\
 & & \sqrt{p} & 0 & 0 & \sqrt{q} & & \\
 & & & & & \ddots & & \\
 & & & &  \sqrt{p} & 0 & 0 & \sqrt{q} \\
 \hline
 & & & & & & \sqrt{q} & * \\
\end{array}\right]_{2(n+1)\times 2(2n+1)}.
\end{equation}
Thus $J_{2n}$ is a special tridiagonal and centrosymmetric matrix as shown below:
\begin{equation}\label{eq:J_2n_matrix}
    J_{2n}=
    \left[
    \begin{array}{ccccc|c}
        0 & \sqrt{p} &  &  &  &  \\
        \sqrt{p} & 0 & \sqrt{pq} & & \\
         & \sqrt{pq} & 0 & \ddots & & \\
        & & \ddots & \ddots & \sqrt{pq} & \\
        & & & \sqrt{pq} & 0 & q \\
        \hline
         & & & & q & *
    \end{array}
    \right]_{2(n+1)\times 2(n+1)}.
\end{equation}

\noindent \textbf{I. eigenvalues of $J_{2n}$}

In order to calculate the characteristic equation $p(\lambda):=|\lambda I -J_{2n}|$ of $J_{2n}$, first introduce the following two principal submatrices.
We use the notation $J_{2n}[l:k]$ to represent the main sub-matrix from $l$ to $k$ rows (columns) of $J_{2n}$, so as to save space.
\begin{align}
    E_k &:= \lambda I - J_{2n}[1:k], \\
    F_k &:= \lambda I -J_{2n}[2:k+1].
\end{align}
Denote by $|M| := \det(M) $ the determinant of matrix $M$. Then we have
\begin{align}
    |E_2| &= \lambda |E_1| - p |E_0|, \label{eq:E_k_recurr_0}\\
    |E_k| &=\lambda |E_{k-1}| -pq |E_{k-2}|,\ (3\leq k \leq n+1) \label{eq:E_k_recurr_1}\\
    |E_k| &=\lambda |F_{k-1}| -p|F_{k-2}|,\ (2\leq k \leq n+1) \label{eq:E_k_recurr_2}\\
    |F_k| &= \lambda |F_{k-1}| -pq |F_{k-2}|,\ (2\leq k \leq n) \label{eq:F_k_recurr}
\end{align}
where the first two terms are $|F_1|=\lambda,\ |F_0|:=1$ and $ |E_1|=\lambda,\ |E_0|:=1$.
Note that Eq.\eqref{eq:E_k_recurr_1} is obtained by expanding $|E_k|$ from its lower right corner, while the second equality is obtained by expanding $|E_k|$ from its upper left corner.
Dividing all the elements in $F_{k}$ by $\sqrt{pq}$ and denoting 
\begin{equation}\label{eq:def_U_k}
    |F_k|/\sqrt{pq}^k := U_k(\lambda/\sqrt{pq}),
\end{equation}
Eq.~\eqref{eq:F_k_recurr} is now transformed to $U_k(\lambda/\sqrt{pq}) =\lambda/\sqrt{pq}\, U_{k-1}(\lambda/\sqrt{pq}) -U_{k-2}(\lambda/\sqrt{pq})$.
If we let $x:=\frac{\lambda}{\sqrt{pq}}$, then Eq.~\eqref{eq:F_k_recurr} further simplifies to
\begin{align}
    U_0(x) &=1, \ U_1(x)=x, \\
    U_{k}(x) &= x\,U_{k-1}(x)-U_{k-2}(x). \label{eq:U_k_def}
\end{align}
Comparing the above equations with the recurrence relation of \textit{Chebyshev polynomial of the second kind}:
\begin{align}
    \tilde{U}_0(x) &=1, \ \tilde{U}_1(x)=2x, \nonumber\\
    \tilde{U}_k(x) &=2x \tilde{U}_{k-1}(x) -\tilde{U}_{k-2}(x), \ \text{for}\ k\geq 2,
\end{align}
we know $U_k(x) = \tilde{U}_k(x/2)$.
From the general term formula $\tilde{U}_k(\cos\theta) = \frac{\sin(k+1)\theta}{\sin\theta}$, we let $\theta$ be such that it satisfies $\frac{\lambda}{2\sqrt{pq}} = x/2 =\cos\theta$. Therefore,
\begin{equation}\label{eq:U_k_cheby}
    U_{k}(\lambda /\sqrt{pq}) = \tilde{U}_k(\cos\theta)=\frac{\sin(k+1)\theta}{\sin\theta},
\end{equation}
from which Eq.~\eqref{eq:U_k_cheby_lem} in Lemma~\ref{lem:spectral_decomp} follows.

We now calculate $p(\lambda) = |\lambda I -J_{2n}|$ by expanding its $n+1$ row as follows. Denote by $E_{k}' = \lambda I -J_{2n}[2n+2-k+1:2n+2]$ the last $k$ rows and columns of $\lambda I -J_{2n}$. It is easy to see that $|E_k'|=|E_k|$ from the centrosymmetry of $J_{2n}$.
\begin{align}
    p(\lambda) &= \sqrt{pq} \left|\begin{array}{ccc} E_{n-1} & & \\ -\sqrt{pq} & -\sqrt{pq} & \\ & -q & E_{n+1}'  \end{array}\right|
    +\lambda \left|\begin{array}{cc} E_n & \\ & E_{n+1} \end{array}\right|
    +q \left|\begin{array}{ccc} E_{n} & -\sqrt{pq} & \\  & -q & -\sqrt{pq} \\ &  & E_{n}'  \end{array}\right| \label{eq:determinant_3}\\
    &= -pq |E_{n-1}|\cdot|E_{n+1}| +\lambda |E_{n}|\cdot|E_{n+1}| -q^2 |E_{n}|^2 \\
    &= (\lambda^2 -q^2) |E_n|^2 -2pq\lambda |E_{n}|\cdot|E_{n-1}| +(pq)^2 |E_{n-1}|^2 \label{eq:expand_E_n_1}\\
    &= \big((\lambda - q) |E_n| -pq |E_{n-1}|\big) \big((\lambda + q) |E_n| -pq |E_{n-1}|\big). \label{eq:two_components} 
\end{align}
Note that when choosing $-q$ as the pivot to expand the first determinant in Eq.~\eqref{eq:determinant_3}, the following sub-determinant is zero.
\begin{equation}
    \left|\begin{array}{c|cc} E_{n-1} & & \\ -\sqrt{pq} & & \\ \hline & -\sqrt{pq} & E_n' \end{array}\right| = 0.
\end{equation}
This is because the first $n$ rows are rank deficient as there are only $n-1$ columns with non zero elements.
The same reasoning applies to expanding the third determinant in Eq.~\eqref{eq:determinant_3}.
The third line (Eq.~\eqref{eq:expand_E_n_1}) uses Eq.~\eqref{eq:E_k_recurr_1} to further expand $|E_{n+1}|$.

We can further simplify the two components of $p(\lambda)$ in Eq.~\eqref{eq:two_components} by using Eqs.~\eqref{eq:E_k_recurr_2},~\eqref{eq:F_k_recurr} to expand until $|F_{n-1}|,\ |F_{n-2}|$:
\begin{align}
    & (\lambda \mp q) |E_n| -pq |E_{n-1}| \label{eq:eigval_1} \\
    &= \lambda |F_{n}| -p|F_{n-1}| \mp q(\lambda |F_{n-1}| -p|F_{n-2}|) \\
    &= \lambda (\lambda |F_{n-1}| -pq |F_{n-2}|) +(\mp (1-p)\lambda -p) |F_{n-1}| \pm pq|F_{n-2}| \\
    &= [\lambda(\lambda\mp 1) \pm p(\lambda \mp 1)]\cdot |F_{n-1}| -pq(\lambda \mp 1) |F_{n-2}| \\
    &= (\lambda \mp 1)\ (|F_{n}| \pm p |F_{n-1}|).
\end{align}
Using Eq.~\eqref{eq:def_U_k}, we now obtain the eigenvalues of $J_{2n}$: $\pm 1$ and $\lambda_{\pm k}:= 2\sqrt{pq} \cos\theta_{\pm k}$, where $\lambda_{\pm k}$ are the $2n$ roots of the following equation:
\begin{equation}\label{eq:eigval_eqn}
\sqrt{q}\, U_n(\lambda/\sqrt{pq}) \pm \sqrt{p}\, U_{n-1}(\lambda/\sqrt{pq}) =0.
\end{equation}
Combining with Eq.~\eqref{eq:U_k_cheby}, we obtain Eq.~\eqref{eq:eigval_eqn_2} in Lemma~\ref{lem:spectral_decomp}.
It can also be seen that when $\theta_{k}$ corresponds to a root with `$+$', then $\theta_{-k} := \pi-\theta_k$ corresponds to a root with `$-$', thus $\lambda_{-k} = -\lambda_k$.

\hspace*{\fill} \\
\noindent \textbf{II. eigenvectors of $J_{2n}$ }

We now consider the eigenvectors of $J_{2n}$.
Using Eqs.~\eqref{eq:E_k_recurr_0}, \eqref{eq:E_k_recurr_1},
and ``$(\lambda_k \mp q) |E_n| -pq |E_{n-1}| = 0$'' by Eq.~\eqref{eq:eigval_1},
it is not difficult to verify that the respective (unnormalized) eigenvector $\ket{v_{\pm k}}$ (which satisfies $(\lambda I -J_{2n}) \ket{v_{\pm k}}=0$) is
\begin{equation}
    [1, \frac{|E_1|}{\sqrt{p}}, \frac{|E_2|}{\sqrt{p}\sqrt{pq}},\cdots, \frac{|E_{n-1}|}{\sqrt{p}\sqrt{pq}^{n-2}}, \frac{|E_n|}{\sqrt{p}\sqrt{pq}^{n-1}}, \pm  (*)],
\end{equation}
Where $(*)$ can be deduced from centrosymmetry.
Combing with the relation of the sub-determinant $|E_k|$, $|F_k|$ shown in Eqs.~\eqref{eq:E_k_recurr_2},~\eqref{eq:F_k_recurr}, the components and the square of the norm of the eigenvector $\ket{v_{\pm k}}$ can be calculated as follows.

\noindent\textbf{1. $\lambda =\pm 1$}

Since $\lambda^2 =(\pm)^2 =1$, we have
\begin{align}
    |F_k| &=\pm |F_{k-1}| -pq |F_{k-2}| \\
    &= \pm q |F_{k-1}| \pm p(\pm |F_{k-2}|-pq|F_{k-3}|) -pq |F_{k-2}| \\
    &= \pm q (|F_{k-1}| -p^2|F_{k-3}|) -p^2 |F_{k-2}|.
\end{align}
Thus we know the difference between every other term, i.e. $|F_k|-p^2 |F_{k-2}|$, is a power series. Therefore,
\begin{align}
    |E_k| &= \pm |F_{k-1}| -p |F_{k-2}|\\
    &= \pm |F_{k}| -p^2 |F_{k-1}| \\
    &= (\pm q)^{k-2} (|F_2| -p^2 |F_0|) \\
    &= (\pm q)^{k-2}(1-pq -p^2) \\
    &= (\pm)^k q^{k-1}.
\end{align}
The second line follows from substituting $q=1-p$ into the second equality of Eq.~\eqref{eq:F_k_recurr}.
Thus the $(i+1)$-th component of $\ket{v_{\pm 1}}$ is
\begin{align}
    \braket{i|v_{\pm 1}} &=\frac{|E_i|}{\sqrt{p}\sqrt{pq}^{i-1}} \\
    &= \frac{(\pm)^i q^{i-1}}{\sqrt{p}\sqrt{pq}^{i-1}} \\
    &= (\pm)^i (\sqrt{{q}/{p}})^i /\sqrt{q},
\end{align}
which is Eq.~\eqref{eq:v_1_element} in Lemma~\ref{lem:spectral_decomp}.
Hence, the square of the norm of $\ket{v_{\pm 1}}$ is:
\begin{align}
    \|\ket{v_{\pm1}}\|^2 &=2(1 +\frac{1}{q} \sum_{i=1}^{n} (\frac{q}{p})^i ) \\
    &=2(1 + \frac{1}{p} \frac{1-(q/p)^n}{1-q/p}) \\
    &= \frac{2}{q-p}( (q/p)^n -2p ),
\end{align}
which is Eq.~\eqref{eq:v_1_norm} in Lemma~\ref{lem:spectral_decomp}.

\hspace*{\fill} \\
\noindent\textbf{2. $\lambda_{\pm k}$ for $k=2\sim (n+1)$}

The $(i+1)$-th component of $\ket{v_{\pm k}}$ can be calculated as follows, where we omit the subscript $\pm k$ for simplicity.
\begin{align}
    \braket{i|v_\lambda} =& \frac{\lambda |F_{i-1}| -p |F_{i-2}|}{\sqrt{p} \sqrt{pq}^{i-1}} \\
    &=\frac{\lambda \sqrt{pq}^{i-1} U_{i-1}(\lambda/\sqrt{pq}) -p \sqrt{pq}^{i-2} U_{i-2}(\lambda/\sqrt{pq})}
    {\sqrt{p} \sqrt{pq}^{i-1}} \\
    &= \frac{\lambda}{\sqrt{p}} U_{i-1}(\lambda/\sqrt{pq}) -\frac{1}{\sqrt{q}} U_{i-2}(\lambda/\sqrt{pq}),
\end{align}
which is Eq.~\eqref{eq:v_k_element} in Lemma~\ref{lem:spectral_decomp}.
We now consider $\| \ket{v_\lambda} \|^2$. We first calculate the square of the $(i+1)$-th component. For simplicity we denote $U_{i}(\lambda/\sqrt{pq})$ by $U_i$ in the following.
\begin{align}
    |\braket{i|v_\lambda}|^2 &= \frac{\lambda^2}{p} U_{i-1}^2 -\frac{2\lambda}{\sqrt{pq}} U_{i-1} U_{i-2} +\frac{1}{q} U_{i-2}^2 \\
    &= \frac{\lambda^2}{p} U_{i-1}^2 +(U_i^2 -\frac{\lambda^2}{pq}U_{i-1}^2 -U_{i-2}^2) +\frac{1}{q} U_{i-2}^2 \\
    &= \frac{p}{q}U_{i-2}^2 -\frac{\lambda^2}{q} U_{i-1}^2 +U_i^2.
\end{align}
The second line is obtained by squaring both sides of the relation $U_i = \frac{\lambda}{\sqrt{pq}} U_{i-1} -U_{i-2}$ which follows from Eq.~\eqref{eq:U_k_def}.
In order to make the relation true for $i\geq 1$, we set $U_{-1}:=0$.
Then, using the identity $p/q -\lambda^2/q +1 =(1-\lambda^2)/q$, we have
\begin{align}
    \| \ket{v_\lambda} \|^2/2 &= 1 +\sum_{i=1}^n |\braket{i|v_\lambda}|^2 \\
    &= 1+ 0 +\frac{p}{q} -\frac{\lambda^2}{q} + \frac{1-\lambda^2}{q} \sum_{i=1}^{n-2}U_i^2 -\frac{\lambda^2}{q}U_{n-1}^2 +U_{n-1}^2 + U_{n}^2 \\
    &=\frac{1-\lambda^2}{q}  +\frac{1-\lambda^2}{q} \sum_{i=1}^{n-2}U_i^2 +(-\frac{\lambda^2}{q}+1+\frac{p}{q})U_{n-1}^2 \\
    &= \frac{1-\lambda^2}{q} \sum_{i=0}^{n-1} U_i^2.
\end{align}
The third line uses Eq.~\eqref{eq:eigval_eqn} satisfied by the eigenvalue $\lambda$.
From the trigonometric expression of $U_i$ (Eq.~\eqref{eq:U_k_cheby}), we have
\begin{align}
\sum_{i=0}^{n-1} U_i^2 &= \frac{1}{\sin^2\theta} \sum_{i=0}^{n-1} \sin^2(i+1)\theta \\
&=\frac{1}{\sin^2\theta} \sum_{i=1}^{n} \frac{1-\cos i2\theta}{2} \\
&= \frac{1}{2\sin^2\theta} (n- \frac{\sin n\theta \cdot \cos(n+1)\theta}{\sin\theta}) \\
&= \frac{1}{2\sin^2\theta} (n \pm \sqrt{\frac{q}{p}} \frac{\sin2(n+1)\theta}{2\sin\theta}).
\end{align}
Hence combined with $\|\ket{u}\|^2 =2(1-\lambda^2)$ in Lemma~\ref{lem:U_convert} and the identity $\theta_{-k} =\pi-\theta_k$, we obtain Eq.~\eqref{eq:v_k_norm} in Lemma~\ref{lem:spectral_decomp}.
Note that the last line above uses Eq.~\eqref{eq:eigval_eqn_2} satisfied by $\theta$.
The third line uses the identity obtained from comparing the real part of the following identities:
\begin{align}
    &\sum_{k=1}^n \cos kx +i\sin kx \\
    &= \sum_{k=1}^n e^{ikx}
    = \frac{e^{ix}(e^{inx}-1)}{e^{ix}-1} \\
    &= \frac{e^{ix} e^{\frac{inx}{2}} 2i \sin\frac{nx}{2}}{e^{\frac{ix}{2}} 2i\sin\frac{x}{2}}
    = \frac{\sin\frac{nx}{2}} {\sin\frac{x}{2}} e^{i\frac{(n+1)x}{2}} \\
    &= \frac{\sin\frac{nx}{2}} {\sin\frac{x}{2}} (\cos\frac{(n+1)x}{2} +i\sin\frac{(n+1)x}{2}).
\end{align}

\section*{Acknowledgements}
We would like to thank Yongzhen Xu and Qingwen Wang for helpful discussions on quantum walk search frameworks.
We would also like to thank the reviewers for suggestions which have improved the presentation of the paper.

\bibliographystyle{quantum}
\bibliography{ref}

\begin{thebibliography}{10}

\bibitem{AharonovDZ93PhysRevA}
Y.~Aharonov, L.~Davidovich, and N.~Zagury.
\newblock ``Quantum random walks''.
\newblock \href{https://dx.doi.org/10.1103/PhysRevA.48.1687}{Phys. Rev. A {\bf 48}, 1687--1690}~(1993).

\bibitem{Kempe_overview}
J.~Kempe.
\newblock ``Quantum random walks: An introductory overview''.
\newblock \href{https://dx.doi.org/10.1080/00107151031000110776}{Contemporary Physics {\bf 44}, 307--327}~(2003).

\bibitem{ambainis_overview}
Andris Ambainis.
\newblock ``Quantum walks and their algorithmic applications''.
\newblock \href{https://dx.doi.org/10.1142/S0219749903000383}{International Journal of Quantum Information {\bf 01}, 507--518}~(2003).

\bibitem{Venegas-Andraca2012}
Salvador~El{\'i}as Venegas-Andraca.
\newblock ``Quantum walks: a comprehensive review''.
\newblock \href{https://dx.doi.org/10.1007/s11128-012-0432-5}{Quantum Information Processing {\bf 11}, 1015--1106}~(2012).

\bibitem{systematic}
Karuna Kadian, Sunita Garhwal, and Ajay Kumar.
\newblock ``Quantum walk and its application domains: A systematic review''.
\newblock \href{https://dx.doi.org/https://doi.org/10.1016/j.cosrev.2021.100419}{Computer Science Review {\bf 41}, 100419}~(2021).

\bibitem{AmbainisBNVW01}
Andris Ambainis, Eric Bach, Ashwin Nayak, Ashvin Vishwanath, and John Watrous.
\newblock ``One-dimensional quantum walks''.
\newblock In Proceedings of the 33rd ACM Symposium on Theory of Computing.
\newblock \href{https://dx.doi.org/10.1145/380752.380757}{Pages 37--49}.
\newblock ~(2001).

\bibitem{AmbainisKV01}
Dorit Aharonov, Andris Ambainis, Julia Kempe, and Umesh~V. Vazirani.
\newblock ``Quantum walks on graphs''.
\newblock In Proceedings of the 33rd ACM Symposium on Theory of Computing.
\newblock \href{https://dx.doi.org/10.1145/380752.380758}{Pages 50--59}.
\newblock ~(2001).

\bibitem{Szegedy_03}
M.~Szegedy.
\newblock ``Quantum speed-up of markov chain based algorithms''.
\newblock In 45th Annual IEEE Symposium on Foundations of Computer Science.
\newblock \href{https://dx.doi.org/10.1109/FOCS.2004.53}{Pages 32--41}.
\newblock ~(2004).

\bibitem{MagniezNRS11}
Fr{\'{e}}d{\'{e}}ric Magniez, Ashwin Nayak, J{\'{e}}r{\'{e}}mie Roland, and Miklos Santha.
\newblock ``Search via quantum walk''.
\newblock \href{https://dx.doi.org/10.1137/090745854}{{SIAM} J. Comput. {\bf 40}, 142--164}~(2011).

\bibitem{KroviMOR16}
Hari Krovi, Fr{\'{e}}d{\'{e}}ric Magniez, Maris Ozols, and J{\'{e}}r{\'{e}}mie Roland.
\newblock ``Quantum walks can find a marked element on any graph''.
\newblock \href{https://dx.doi.org/10.1007/s00453-015-9979-8}{Algorithmica {\bf 74}, 851--907}~(2016).

\bibitem{belovs2013quantum}
Aleksandrs Belovs.
\newblock ``Quantum walks and electric networks''~(2013).
\newblock  \href{http://arxiv.org/abs/1302.3143}{arXiv:1302.3143}.

\bibitem{unified}
Simon Apers, Andr\'{a}s Gily\'{e}n, and Stacey Jeffery.
\newblock ``{A Unified Framework of Quantum Walk Search}''.
\newblock In Markus Bl\"{a}ser and Benjamin Monmege, editors, 38th International Symposium on Theoretical Aspects of Computer Science (STACS 2021).
\newblock \href{https://dx.doi.org/10.4230/LIPIcs.STACS.2021.6}{Volume 187 of Leibniz International Proceedings in Informatics (LIPIcs), pages 6:1--6:13}.
\newblock Dagstuhl, Germany~(2021). Schloss Dagstuhl -- Leibniz-Zentrum f{\"u}r Informatik.

\bibitem{Ambainis07}
Andris Ambainis.
\newblock ``Quantum walk algorithm for element distinctness''.
\newblock \href{https://dx.doi.org/10.1137/S0097539705447311}{{SIAM} J. Comput. {\bf 37}, 210--239}~(2007).

\bibitem{BuhrmanS06}
Harry Buhrman and Robert Spalek.
\newblock ``Quantum verification of matrix products''.
\newblock In Proceedings of the Seventeenth ACM-SIAM Symposium on Discrete Algorithms.
\newblock Pages 880--889.
\newblock ~(2006).
\newblock  url:~\url{http://dl.acm.org/citation.cfm?id=1109557.1109654}.

\bibitem{MagniezSS07}
Fr{\'{e}}d{\'{e}}ric Magniez, Miklos Santha, and Mario Szegedy.
\newblock ``Quantum algorithms for the triangle problem''.
\newblock \href{https://dx.doi.org/10.1137/050643684}{{SIAM} J. Comput. {\bf 37}, 413--424}~(2007).

\bibitem{MagniezN07}
Fr{\'{e}}d{\'{e}}ric Magniez and Ashwin Nayak.
\newblock ``Quantum complexity of testing group commutativity''.
\newblock \href{https://dx.doi.org/10.1007/s00453-007-0057-8}{Algorithmica {\bf 48}, 221--232}~(2007).

\bibitem{CCD03}
Andrew~M. Childs, Richard Cleve, Enrico Deotto, Edward Farhi, Sam Gutmann, and Daniel~A. Spielman.
\newblock ``Exponential algorithmic speedup by a quantum walk''.
\newblock In Proceedings of the Thirty-Fifth Annual ACM Symposium on Theory of Computing.
\newblock \href{https://dx.doi.org/10.1145/780542.780552}{Page 59–68}.
\newblock STOC '03New York, NY, USA~(2003). Association for Computing Machinery.

\bibitem{childs2002example}
Andrew~M. Childs, Edward Farhi, and Sam Gutmann.
\newblock ``An example of the difference between quantum and classical random walks''.
\newblock \href{https://dx.doi.org/10.1023/A:1019609420309}{Quantum Information Processing {\bf 1}, 35--43}~(2002).

\bibitem{kempe2002quantum}
Julia Kempe.
\newblock ``Quantum random walks hit exponentially faster''~(2002).
\newblock  \href{http://arxiv.org/abs/quant-ph/0205083}{arXiv:quant-ph/0205083}.

\bibitem{multi}
Stacey Jeffery and Sebastian Zur.
\newblock ``Multidimensional quantum walks''.
\newblock In Proceedings of the 55th Annual ACM Symposium on Theory of Computing.
\newblock \href{https://dx.doi.org/10.1145/3564246.3585158}{Page 1125–1130}.
\newblock STOC 2023New York, NY, USA~(2023). Association for Computing Machinery.

\bibitem{childs2020can}
Andrew~M. Childs and Daochen Wang.
\newblock ``Can graph properties have exponential quantum speedup?''~(2020).
\newblock  \href{http://arxiv.org/abs/2001.10520}{arXiv:2001.10520}.

\bibitem{Ben_2020}
Shalev Ben-David, Andrew~M. Childs, András Gilyén, William Kretschmer, Supartha Podder, and Daochen Wang.
\newblock ``Symmetries, graph properties, and quantum speedups''.
\newblock In 2020 IEEE 61st Annual Symposium on Foundations of Computer Science (FOCS).
\newblock \href{https://dx.doi.org/10.1109/FOCS46700.2020.00066}{Pages 649--660}.
\newblock ~(2020).

\bibitem{exact_simon}
G.~Brassard and P.~Hoyer.
\newblock ``An exact quantum polynomial-time algorithm for simon's problem''.
\newblock In Proceedings of the Fifth Israeli Symposium on Theory of Computing and Systems.
\newblock \href{https://dx.doi.org/10.1109/ISTCS.1997.595153}{Pages 12--23}.
\newblock ~(1997).

\bibitem{GSP}
Zekun Ye, Yunqi Huang, Lvzhou Li, and Yuyi Wang.
\newblock ``Query complexity of generalized simon's problem''.
\newblock \href{https://dx.doi.org/https://doi.org/10.1016/j.ic.2021.104790}{Information and Computation {\bf 281}, 104790}~(2021).

\bibitem{localized}
Yusuke Ide, Norio Konno, Etsuo Segawa, and Xin-Ping Xu.
\newblock ``Localization of discrete time quantum walks on the glued trees''.
\newblock \href{https://dx.doi.org/10.3390/e16031501}{Entropy {\bf 16}, 1501--1514}~(2014).

\bibitem{upper_improve}
Yosi Atia and Shantanav Chakraborty.
\newblock ``Improved upper bounds for the hitting times of quantum walks''.
\newblock \href{https://dx.doi.org/10.1103/PhysRevA.104.032215}{Phys. Rev. A {\bf 104}, 032215}~(2021).

\bibitem{QSVT}
Andr\'{a}s Gily\'{e}n, Yuan Su, Guang~Hao Low, and Nathan Wiebe.
\newblock ``Quantum singular value transformation and beyond: Exponential improvements for quantum matrix arithmetics''.
\newblock In Proceedings of the 51st Annual ACM SIGACT Symposium on Theory of Computing.
\newblock \href{https://dx.doi.org/10.1145/3313276.3316366}{Page 193–204}.
\newblock STOC 2019New York, NY, USA~(2019). Association for Computing Machinery.

\bibitem{fixed_point}
Theodore~J. Yoder, Guang~Hao Low, and Isaac~L. Chuang.
\newblock ``Fixed-point quantum search with an optimal number of queries''.
\newblock \href{https://dx.doi.org/10.1103/PhysRevLett.113.210501}{Phys. Rev. Lett. {\bf 113}, 210501}~(2014).

\bibitem{lower_improve}
Stephen~A. Fenner and Yong Zhang.
\newblock ``A note on the classical lower bound for a quantum walk algorithm''~(2003).
\newblock  \href{http://arxiv.org/abs/quant-ph/0312230}{arXiv:quant-ph/0312230}.

\bibitem{childs2010relationship}
Andrew~M. Childs.
\newblock ``On the relationship between continuous- and discrete-time quantum walk''.
\newblock \href{https://dx.doi.org/10.1007/s00220-009-0930-1}{Communications in Mathematical Physics {\bf 294}, 581--603}~(2010).

\bibitem{phase_estimation}
A.~Yu. Kitaev.
\newblock ``Quantum measurements and the abelian stabilizer problem''~(1995).
\newblock  \href{http://arxiv.org/abs/quant-ph/9511026}{arXiv:quant-ph/9511026}.

\bibitem{BV}
Ethan Bernstein and Umesh Vazirani.
\newblock ``Quantum complexity theory''.
\newblock In Proceedings of the Twenty-Fifth Annual ACM Symposium on Theory of Computing.
\newblock \href{https://dx.doi.org/10.1145/167088.167097}{Page 11–20}.
\newblock STOC '93New York, NY, USA~(1993). Association for Computing Machinery.

\bibitem{exact}
Guanzhong Li and Lvzhou Li.
\newblock ``Deterministic quantum search with adjustable parameters: Implementations and applications''.
\newblock \href{https://dx.doi.org/https://doi.org/10.1016/j.ic.2023.105042}{Information and Computation {\bf 292}, 105042}~(2023).

\bibitem{Long}
G.~L. Long.
\newblock ``Grover algorithm with zero theoretical failure rate''.
\newblock \href{https://dx.doi.org/10.1103/PhysRevA.64.022307}{Phys. Rev. A {\bf 64}, 022307}~(2001).

\bibitem{Grover}
Lov~K. Grover.
\newblock ``A fast quantum mechanical algorithm for database search''.
\newblock In Proceedings of the Twenty-Eighth Annual ACM Symposium on Theory of Computing.
\newblock \href{https://dx.doi.org/10.1145/237814.237866}{Page 212–219}.
\newblock STOC '96New York, NY, USA~(1996). Association for Computing Machinery.

\bibitem{video}
``Evolution of the reduced state vector of {DTQW} on welded tree''.
\newblock \url{https://www.bilibili.com/video/BV1kK411179r/}.

\bibitem{Jordan_1875}
Camille Jordan.
\newblock ``Essai sur la géométrie à $n$ dimensions''.
\newblock Bulletin de la Société Mathématique de France {\bf 3}, 103--174~(1875).
\newblock  url:~\url{http://eudml.org/doc/85325}.

\bibitem{eedp}
Guanzhong Li and Lvzhou Li.
\newblock ``Optimal exact quantum algorithm for the promised element distinctness problem''~(2022).
\newblock  \href{http://arxiv.org/abs/2211.05443}{arXiv:2211.05443}.

\end{thebibliography}
\end{document}